\documentclass{aa}
\usepackage{graphicx}
\usepackage{natbib}
\usepackage{txfonts}
\bibpunct{(}{)}{;}{a}{}{,}
\newcommand{\Ds}{D_{\mathrm{s}}}

\newcommand{\Dls}{D_{\mathrm{ls}}}
\newcommand{\D}{\mathrm{D}}
\newcommand{\zl}{z_{\mathrm{l}}}

\newcommand{\zs}{z_{\mathrm{s}}}
\newcommand{\zr}{z_{\mathrm{r}}}

\newcommand{\myarcsec}{\hbox{$.\!\!^{\prime\prime}$}}
\newcommand{\myarcmin}{\hbox{$.\!\!^{\prime}$}}
\newcommand{\myarcsecnodot}{\hbox{$\;\!\!^{\prime\prime}\;$}}
\newcommand{\myarcminnodot}{\hbox{$^{\prime}\;$}}

\newcommand{\angstromblank}{\AA$\;$}

\def\cm3{\rm ~cm^{-3}}
\def\kms{\rm ~km~s^{-1}}

\def\ergs{\rm ~erg~s^{-1}}

\begin{document}
 
\bibliographystyle{aa}

\title{J0454-0309: Evidence of a strong lensing fossil group\\
       falling into a poor galaxy cluster
\thanks{This work is based on data collected at the Subaru
  Telescope, which is operated by the National Astronomical
  Observatory of Japan; based on observations obtained 
  with MegaPrime/MegaCam, a joint project of CFHT and
  CEA/DAPNIA, at the Canada-France-Hawaii Telescope (CFHT) which
  is operated by the National Research Council (NRC) of Canada,
  the Institut National des Sciences de l'Univers of the Centre
  National de la Recherche Scientifique (CNRS) of France, and the
  University of Hawaii; based on observations made with ESO
  Telescopes at the La Silla and Paranal Observatories, 
  Chile (ESO DDT Programme 282.A-5066); based on observations
  made with the NASA/ESA Hubble Space Telescope (programme \#9836)
  obtained at the Space Telescope Science Institute, which is
  operated by the Association of Universities for Research in
  Astronomy, Inc., under NASA contract NAS 5-26555; based on
  observations obtained with XMM-Newton, an ESA science mission
  with instruments and contributions directly funded by ESA
  Member States and NASA; based on data obtained at the W.M. Keck
  Observatory, which is operated as a scientific partnership among the
  California Institute of Technology, the University of California and
  the National Aeronautics and Space Administration. The 
  Observatory was made possible by the generous financial support of
  the W.M. Keck Foundation.}}

\author{Mischa Schirmer\inst{1}
      \and
        Sherry Suyu\inst{1}
      \and
        Tim Schrabback\inst{2}
       \and
        Hendrik Hildebrandt\inst{2}
       \and
        Thomas Erben\inst{1}
       \and
        Aleksi Halkola\inst{3,4}
       }

\offprints{mischa@astro.uni-bonn.de}

\institute{Argelander-Institut f\"ur Astronomie, Universit\"at
  Bonn, Auf dem H\"ugel 71, 53121 Bonn, Germany
  \and
  Leiden Observatory, Leiden University, Niels Bohrweg 2, 2333 Leiden, The Netherlands
  \and
  University of Tuorla Observatory, V\"ais\"al\"antie 20, 21500
  Piikki\"o, Finland
  \and
  Excellence Cluster Universe, Technische Universit\"at M\"unchen,
  Boltzmannstr. 2, 85748 Garching, Germany
}

\date{Received 5 December 2009; accepted 4 February 2010}

\abstract 
{}
{We have discovered a strong lensing fossil group (J0454) 
  projected near the well-studied cluster MS0451-0305. Using the large 
  amount of available archival data, we compare J0454 to normal groups 
  and clusters. A highly asymmetric image configuration of the strong
  lens enables us to study the substructure of the system.}
{We used multicolour Subaru/Suprime-Cam and CFHT/Megaprime
  imaging, together with Keck spectroscopy to identify member galaxies.
  A VLT/FORS2 spectrum was taken to determine the redshifts of the
  brightest elliptical and the lensed arc. Using HST/ACS images, we
  determined the group's weak lensing signal and modelled the strong lens
  system. This is the first time that a fossil group is analysed with
  lensing methods. The X-ray luminosity and temperature were derived
  from XMM-Newton data.}
{J0454 is located at $z=0.26$, with a gap of 2.5 mag between the
  brightest and second brightest galaxies within half 
  the virial radius. Outside a radius of 1.5 Mpc we find two
  filaments extending over $4$ Mpc, and within we identify 31 members 
  spectroscopically and 33 via the red sequence with $i<22$ mag. 
  They segregate into spirals ($\sigma_v=590\,\kms$) and a central 
  concentration of ellipticals ($\sigma_v=480\,\kms$), establishing
  a morphology-density relation. Weak lensing and cluster richness
  relations yield consistent values of $r_{200}=810-850$ kpc and 
  $M_{200}=(0.75-0.90)\times10^{14}{\rm M_\odot}$. The brightest group
  galaxy (BGG) is inconsistent with the dynamic centre of J0454. It
  strongly lenses a galaxy at $z=2.1\pm0.3$, and we model the lens
  with a pseudo-isothermal elliptical mass distribution. A high
  external shear, and a discrepancy between the Einstein radius and
  the weak lensing velocity dispersion requires that the BGG must be
  offset from J0454's dark halo centre by at least $90-130$ kpc. The
  X-ray halo is offset by $24\pm16$ kpc from the BGG, shows no signs
  of a cooling flow and can be fit by a single $\beta$-model. With
  $L_X=(1.4\pm0.2)\times10^{43}\ergs$ J0454 falls onto standard
  cluster scaling relations, but appears cooler ($T=1.1\pm0.1$ keV)
  than expected ($T\sim2.0$ keV). Taken all together, these data
  indicate that J0454 consists of two systems, a sparse cluster and an
  infalling fossil group, where the latter seeds the brightest cluster
  galaxy. An alternative to the sparse cluster could be a filament
  projected along the line of sight mimicking a cluster, with
  galaxies streaming towards the fossil group.}  
{}

\keywords{Galaxies: clusters: individual: J0454-0309, Galaxies:
  formation, Galaxies: evolution, Gravitational lensing: strong,
  Gravitational lensing: weak}

\titlerunning{The fossil group J0454-0309}

\authorrunning{Schirmer et al.}

\maketitle

\section{\label{intro}Introduction}
In a $\Lambda$CDM cosmology galaxies acquire mass mostly through
minor merger events, where one galaxy has 0.3 times or less the 
mass of its collision partner. Only the most luminous elliptical
galaxies experience a major merger event in their history
\citep{pef09}. The growth of large elliptical systems is
facilitated particularly well in the low-velocity environments of galaxy
groups where dynamical friction \citep{cha43,nus99} is very efficient.
This effect increases with the mass of the infalling galaxy and is
higher for lower velocities. In this way galaxies cool down into the
group or cluster core, losing their gas through ram-pressure stripping
along the way \citep{qmb00}. At the same time they undergo slower
morphological transformations \citep{pef09,sws09}, leading to the
formation of the red sequence in the inner region. The time
scale for dynamical friction depends on 
the mass of the infalling galaxy and its distance from the core. For
the most massive galaxies ($\sim 10^{12}{\rm M_\odot}$) it is as short as a
few Gyrs \citep{nat08,bmq08}, implying that large elliptical galaxies
in groups can already accur at early times.

Several mechanisms for the formation of BCGs (the brightest cluster 
galaxies) have been suggested, ranging from galactic cannibalism and
cooling flows to merger processes during cluster collapse
\citep[see][and references therein]{lbk07}. Elliptical galaxies
growing in this fashion should be located at the centre of the
gravitational potential, and their recession velocity should match the
mean of the radial velocities of the other cluster members for
virialised systems. Recently, \cite{sby10} have shown that in
$\sim40\%$ of all haloes of mass $\sim5\times10^{13}\,h_{100}\,{\rm M_\odot}$
the BCG is not the central galaxy, falsifying this paradigm. This was
also demonstrated for clusters with higher masses
\cite[][]{oeh01,lbk07}. Most of these analyses have in
common that the centre of the halo is identified by the distribution 
centre of elliptical galaxies or, more rarely, by the X-ray centroid
or weak gravitational lensing. Either of these methods has advantages
and disadvantages, for instance, they can be hampered by small numbers
of galaxies, low X-ray S/N or projection effects. In this paper we are 
in the lucky situation that a strongly lensed galaxy let us put tight 
constraints on the dark matter halo centre, and in this way show
that the BCG is not located at the minimum of the
potential. We conclude that the BCG was formed outside the cluster in
a nearby group, which is now falling into the cluster.

\subsection{Fossil groups}
Contrary to the quick dynamical collapse of galaxy groups, the
cooling times for their X-ray haloes are comparable to 
one or several Hubble times \citep{sar88}. This can lead to isolated
giant elliptical galaxies, embedded in X-ray haloes with luminosities
characteristic for entire galaxy groups. Such objects exist in
the Universe \citep{vnh99}, either isolated \citep{yft04} or
surrounded by groups of less luminous satellite galaxies
\citep{jph03,kpj06,bcr09}. One of the first systems has been reported
by \cite{paj94} coining the term `fossil group', and \cite{jph03}
have introduced general selection criteria. Accordingly, the galaxies
must be embedded in an extended X-ray halo with 
$L_X>10^{42}\,h_{50}^{-2}\ergs$, integrated over the $0.5-2.0$ keV
range. In addition, the central elliptical galaxy must be 
$\Delta m_{12}^{\rm min}\geq 2$ mag brighter in $R$-band than the
second brightest galaxy (independent of morphology) within half
the virial radius. This magnitude gap is motivated by the accretion of
$L_*$-galaxies in the inner volume, which are then absent in the
group's luminosity function. Current observational samples 
\citep[e.g.][]{kpj07,bcr09,vbh10} are largely based on this
definition, and so are simulations \citep{bog08}.

The selection criteria by \cite{jph03} have been relaxed in the course
of systematic searches. \cite{sms07} have favoured a fixed
radius of $0.5\,h_{50}$ Mpc within which the magnitude 
gap must hold, independent of the cluster's virial state. \cite{vbh10}
adopt $0.7\,r_{500}\sim0.4\,r_{\rm vir}$, with $r_{500}$ being
calculated from the group's X-ray luminosity. Similar relaxations have 
been adopted for the magnitude gap. \cite{mmf06} and \cite{bcr09} show
that there is no sharp transition in the magnitude gap of galaxy
clusters, hence there is no physical motivation for a particular
numeric value. \cite{vbh10} and \cite{bcr09} favour smaller gap sizes
of $\Delta m_{12}^{\rm min}=1.7$ and 1.75 mag, respectively. According
to \cite{vbh10} the gap should not be too strict a requirement, as the
determination of the total magnitude of very extended galaxies is not
trivial.

As for the formation of the magnitude gap, \cite{bog08} have found in
simulations that it usually arises at redshifts $0<z<0.7$, after the
haloes assembled half of their final mass at $0.8<z<1.2$. This is
significantly earlier than the formation of normal groups
\citep{dkp07} and leads to increased NFW \citep{nfw97} concentration
parameters. Accordingly, the last major merger in the simulated fossil
groups took place more than 6 Gyrs ago for more than 50\% of the
galaxies. Most of the magnitude gaps are closed at later (current)
times when more infall of satellite galaxies occurs.

The exact formation process of fossils is not yet entirely understood.
For example, \cite{yft04} observe mass-to-light ratios as high as
$1000$, which are difficult to explain if these galaxies assembled
their mass only through dynamical friction. Another uncertainty lies
in the type of the galaxies from which the giant elliptical forms. 
\cite{kpj06} argue that their disky isophotes indicate gas-rich
mergers, which would distinguish these galaxies from the BCGs in
normal clusters. These tend to show more boxy isophotes from gas-poor
mergers. However, \cite{bcr09} do not find a preference for either
disky or boxy shapes in their larger sample. They have argued
that fossils merely represent a transitional state in the
last stages of mass assembly than a class of their own.

While the formation process of fossils is still a matter of debate,
their occurrence is not. About $10\%-20\%$ of all X-ray luminous
groups and clusters have fossil character \citep[e.g.][]{jph03,bog08},
with typical masses of $1-10\times10^{13} {\rm M_\odot}$. However, they are
difficult to identify observationally. Only a few dozen systems
are known so far, mostly extracted from large-area
surveys such as SDSS \citep{sms07,bcr09} or the 400D cluster catalogue
\citep{vbh10}. The last authors discuss various difficulties in the
selection process, in particular completeness and
problems in the accurate determination of the magnitude of the
brightest galaxy. In general, the observationally determined
abundances of fossil groups agree with those predicted by
simulations. However, in terms of absolute numbers samples are
systematically incomplete since the second brightest galaxy can be at
a sufficiently large physical distance from the centre and still
appear projected onto the inner volume. Assuming that all galaxies
are within the virial radius and follow a radially symmetric
distribution, we estimate that $20\%$ $(25\%)$ of all fossils
are overlooked for $r_{\rm min}=0.4$ (0.5) 
$r_{\rm vir}$ due to this effect.

Almost all of the few dozen fossil groups known were discovered and
analysed based upon comparatively shallow optical and/or X-ray survey
data. In general the observational data are poor compared to what are
available for normal clusters. Only a few fossils were investigated in
detail, such as ESO 3060170 \citep{sfv04}, RXJ1552.2+2013
\citep{mcs06}, RXJ1416.4+2315 \citep{jph03,cms06,kmp06}, CL0259+0013
\citep{vbh10} or UGC 842 \citep{lcm10}. For a comprehensive comparison
with normal groups and clusters a systematic deep survey of a larger
number is needed.

\begin{figure}[t]
  \includegraphics[width=1.0\hsize]{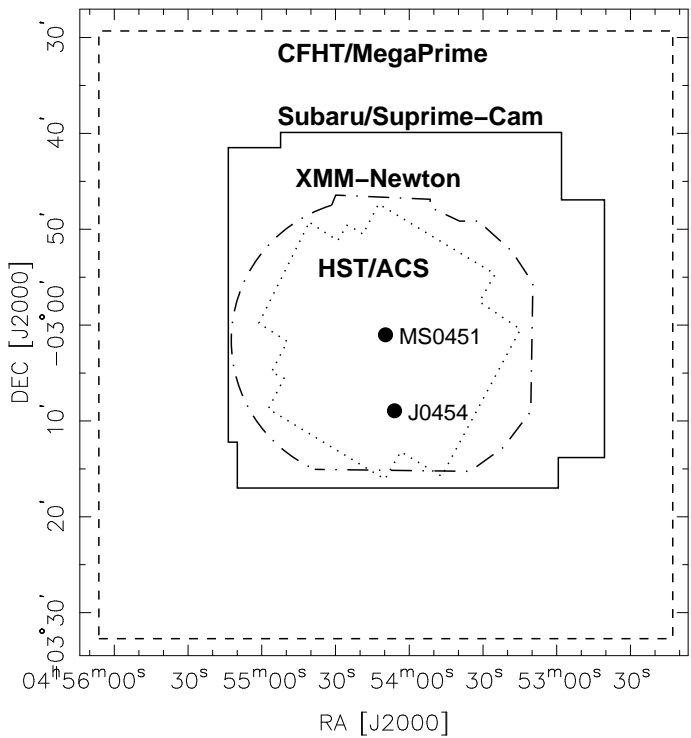}
  \caption{\label{fields}{Pointings of the imaging data sets.
      The positions of the fossil group J0454 and the 
      background cluster MS0451 are shown as well.}}
\end{figure}

In this paper we present our analysis of J0454.0-0308 (hereafter: 
J0454), a fossil group at $z=0.26$. It is projected 
8\myarcminnodot south of the well-known cluster MS0451-0301
(hereafter: MS0451, $z=0.54$), thus a large amount of
archival data are available for our analysis. J0454 consists of at
least 60 galaxies and was identified by us in Subaru/Suprime-Cam
images. It is dominated by a giant elliptical galaxy (hereafter:
E0454), which strongly lenses a distant background source. We use
Subaru/Suprime-Cam and CFHT/MegaPrime for photometry, XMM-Newton
to study the intra-cluster gas and HST/ACS for the weak and strong
lensing analysis. The imaging data (see Fig. \ref{fields} for an
overview) are complemented by VLT and Keck spectroscopy. 

\subsection{Terminology and assumptions}
In this work we present evidence that J0454 is composed of a poor
cluster and an infalling fossil group. We refer to the global
system as J0454, but also to the cluster without the fossil
group. The latter distinction is only made in Sect. 8 when  
we discuss the results. E0454 is the brightest galaxy of the system.

The paper is organised as follows: In Sect. 2 we describe the imaging
and spectroscopic data provided this was not done elsewhere. In
Sect. 3 we study foreground and background contamination and select a
red cluster sequence in colour-colour and colour-magnitude space. In
Sect. 4 we investigate the galactic content of the system, establish a
morphology-density relation and obtain the velocity dispersions of
early- and late-type galaxies. We use virial properties and the
size-richness relation for an estimate of $r_{200}$. In Sect. 5 we
present the X-ray results, followed by our weak and strong lensing
analysis in Sects. 6 and 7. We discuss our findings in Sect. 8 and
summarise in Sect. 9.

We assume a flat standard cosmology with $\Omega_m=0.27$,
$\Omega_\Lambda=0.73$ and $H_0=70\,h\kms\,{\rm Mpc}^{-1}$. On 
occasion we refer to relations from the literature with
parameterisations $H_0=100\,h_{100}\kms\,{\rm Mpc}^{-1}$ or
$H_0=50\,h_{50}\kms\,{\rm Mpc}^{-1}$. To avoid confusion 
we quote them as published originally, indexing $h$ accordingly. X-ray
luminosities are reported for the $0.5-2.0$ keV range, and optical
luminosities are given in solar units. The relation between angular
and physical scales at $z=0.26$ is $1^{\prime}=243\,h^{-1}$ kpc. All
numeric values quoted for physical distances in J0454 must be scaled
with $h^{-1}$. Magnitudes are reported for both the Johnson-Cousins and
the Sloan passbands and denoted with uppercase and lowercase
letters, respectively. All error bars represent the $1\sigma$
confidence level.

\section{Observations and data reduction}
\subsection{Subaru/Suprime-Cam and CFHT/Megaprime data reduction}
We serendipitously discovered J0454 in deep Subaru/Suprime-Cam
\citep{mks02} images of MS0451. The data were reduced with
THELI\footnote{Available at
http://www.astro.uni-bonn.de/$\sim$mischa/theli.html} \citep{esd05},
our pipeline for the reduction of wide-field optical and near-infrared
images. In the following we summarise those aspects where our
reduction scheme deviated significantly from the standard approach.

Images were taken in nine different nights during six periods
between 2001-01-22 and 2006-12-21 (PIs: H. Ebeling, N. Yasuda,
G. Kosugi). Suprime-Cam consists of 10 CCDs, covering
$34^{\prime}\times27^{\prime}$ with 0\myarcsec202 per pixel. In 2001
Suprime-Cam had one broken CCD and individual gain settings. The
defect CCD and three others were replaced, and the gains were
homogenised and refined once more another year later. We brought all
chips to the same gain and then performed the standard pre-processing
including debiasing, flatfielding, superflatting, defringing, and sky
subtraction. The data were astrometrically calibrated with
\textit{Scamp} \citep{ber06} and then stacked. Since images were taken
with two different sky position angles we could recover areas
initially lost due to blooming. The data did not allow for correction
of scattered light in the flat fields, for which extensive dithering
of photometric standard fields is required 
\citep[see][]{mas01,mac04,kog04}.

We complemented the $BVRIz$ Subaru/Suprime-Cam data with $u^*griz$
CFHT/Megaprime images, which improves the photometric redshifts as a
result of the presence of $u^*$-band. The CFHT/Megaprime data were 
pre-reduced using {\tt ELIXIR} \citep{mac04} at CFHT, including
corrections for scattered light of the order of 0.1 mag. The remaining
processing was done with THELI following \cite{ehl09}. The properties
of the coadded images are summarised in Table \ref{table_data_sup}.

\subsubsection{\label{catcreation}Catalogue creation}
Object detection and photometry was done using SExtractor
\citep{bea96} in double image mode. We stacked all exposures in all
filters of one camera with an image seeing of less than 1\myarcsec0,
obtaining a deep noise-normalised detection image. Coadded images in
the different filters were convolved to a common seeing of
0\myarcsec95, ensuring that the object flux in each waveband was
integrated over identical apertures. We kept objects with at least 5
connected pixels with $S/N\geq 2$ each.

The Subaru/Suprime-Cam data were only partially taken in photometric
conditions, with zeropoint variations of up to 0.1 mag in other
nights. We tied the photometric $z$-band image to CFHT/Megaprime data
taken in the same filter. The other Subaru/Suprime-Cam zeropoints were
inferred by comparing the fluxes from non-saturated stars, measured in
$3^{\prime\prime}$ wide apertures, against the \cite{pic98}
library \citep[for details see][]{ehl09}. We took into account filter
transmission, quantum efficiency, and the combined mirror reflectivity
and corrector throughput (Table \ref{subaru_transmission},
S. Miyazaki, priv. comm.). The photometric calibration of the
CFHT/Megaprime data was taken from the {\tt ELIXIR} headers.
The zeropoints of both data sets were ultimately fine-tuned during the
calculation of the photometric redshifts based on several hundred
calibration spectra (see Sect. \ref{photz}).

\begin{figure}[t]
  \includegraphics[width=1.0\hsize]{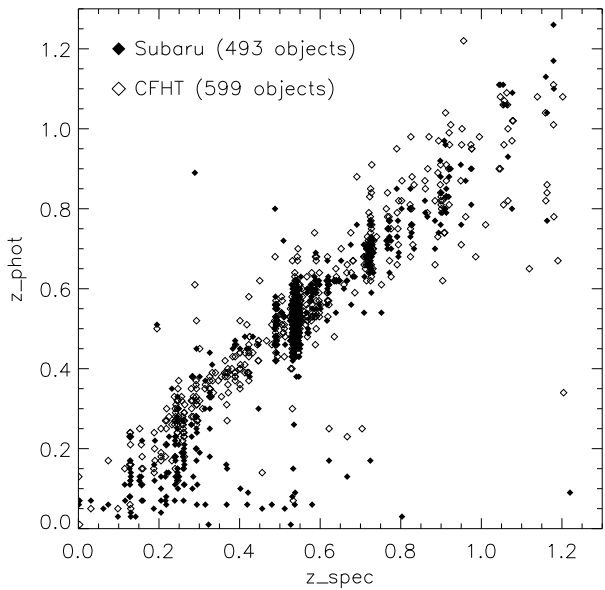}
  \caption{\label{specphotz}{Comparison of photometric and
      spectroscopic redshifts.}}
\end{figure}

\subsubsection{\label{photz}Photometric redshifts}
We need photometric redshifts for the weak gravitational lensing
analysis in Sect. \ref{weaklens}, mainly to distinguish between
lensed background and unlensed foreground galaxies. The photometric
redshifts were obtained as outlined in \cite{hpe09} for all objects in
the catalogues (see Sect. \ref{catcreation}) and calibrated
against 774 and 493 spectroscopic redshifts from \cite{met07}, for
CFHT and Subaru, respectively. Notice that \cite{met07} obtained
spectroscopic redshifts for a total of 1562 sources in the field of
view of MS0451. We performed the phot-z calibration using only those
spectra of sources with photometric errors smaller than 0.1 mag in all
bands. In detail, we fix the
redshifts of the corresponding galaxies to their spectroscopically 
determined values. The magnitude differences between the best-fit 
templates and the observed photometry then yield the zeropoint 
corrections, in the range of $0.02-0.09$ mag for CFHT 
and $0.04-0.18$ mag for Subaru. The correlation between
photometric and spectroscopic redshifts is shown in
Fig. \ref{specphotz} for both data sets, using a confidence limit
(ODDS parameter) higher than $0.8$. Due to the lack of $u$-band data
the Subaru photo-zs are highly unreliable for $z\lesssim0.3$, moving a
significant fraction of lensed galaxies into the unlensed foreground
sample. The CFHT data are much better in this respect, but shows $2-3$
times as much scatter for $z\gtrsim0.6$ and fails for fainter galaxies
due to inferior depth in $i$- and $z$-band. The accuracy of the
photo-zs is $\sigma\sim0.040$.

We run both data sets simultaneously through the photo-z process but
found the results to be significantly worse than the photo-zs obtained
separately for CFHT and Subaru. This is due to different PSF
characteristics of the two data sets which could not be homogenised
sufficiently, and in particular due to the fact that the Subaru data
could not be corrected for scattered light in the flats, resulting in
inconsistent magnitudes across similar passbands. We therefore created
a composite photo-z catalogue in the following manner. We took the CFHT
estimate if $z_{\rm phot}^{\rm CFHT}<=0.4$ and the average if both
estimates are between 0.4 and 0.7. The remaining galaxies were split
in two groups. The first is formed by galaxies for which the Subaru
redshift is higher than 0.7, and we assigned them this
estimate. Galaxies in the second group, with 
$z_{\rm phot}^{\rm Subaru}<=0.7$ and $z_{\rm phot}^{\rm CFHT}>0.4$ got
either the CFHT or the Subaru redshift assigned, depending on which
one has higher confidence. Ultimately, the redshifts were transformed
into relative lensing strengths, 
\begin{equation}
\beta=\frac{\D(\zl,\zs)}{\D(0,\zs)},
\label{beta}
\end{equation}
where $\D(z_1,z_2)$ is the angular diameter distance between two
sources at redshifts $z_1$ and $z_2$, and $\zl$ and $\zs$ are the lens
and source redshifts, respectively. See Sects. \ref{acs} and 
\ref{weaklens} for more details.

\begin{table}
\caption{Summary of the Subaru/Suprime-Cam and CFHT/Megaprime
  data. The limiting AB magnitudes (50\% completeness limit) are for
  $10\sigma$ point sources, and are on average 0.8 mag brighter for
  extended objects.}
\label{table_data_sup}
\begin{tabular}{l r r r r}
\hline
\hline
Telescope/Instrument & Filter (abbr.) & $t_{\rm exp}$ [s] & Seeing & $M_{\rm lim}$\\
\hline
\noalign{\smallskip}
Subaru/Suprime-Cam & WJB  ($B$) & 12240 & 0\myarcsec82 & 26.7 \\
Subaru/Suprime-Cam & WJV  ($V$) & 5040  & 0\myarcsec95 & 26.0 \\
Subaru/Suprime-Cam & WCRC ($R$) & 11400 & 0\myarcsec83 & 26.6 \\
Subaru/Suprime-Cam & WCIC ($I$) & 4920  & 0\myarcsec92 & 25.9 \\
Subaru/Suprime-Cam & WSZ  ($z$) & 4380  & 0\myarcsec76 & 25.1 \\
\hline
CFHT/Megaprime & $u^*$ & 5220  & 0\myarcsec87 & 25.7\\
CFHT/Megaprime & $g$   & 3400  & 0\myarcsec85 & 26.0\\
CFHT/Megaprime & $r$   & 14850 & 0\myarcsec71 & 26.2\\
CFHT/Megaprime & $i$   & 1280  & 0\myarcsec71 & 23.7\\
CFHT/Megaprime & $z$   & 1440  & 0\myarcsec70 & 22.4\\
\hline
\end{tabular}
\end{table}

\begin{table}
\caption{Combined Subaru mirror reflectivity and corrector throughput}
\label{subaru_transmission}
\begin{tabular}{c c}
\hline
\hline
Wavelength [\AA]& Throughput\\
\hline
\noalign{\smallskip}
4450  &  0.774 \\
5500  &  0.828 \\
6590  &  0.828 \\
7710  &  0.791 \\
9220  &  0.765 \\
\hline
\end{tabular}
\end{table}

\subsection{\label{acs}HST/ACS imaging and shear catalogue} 
For the weak lensing measurements and the strong lens modelling we
rely on wide-field imaging with HST/ACS through the F814W filter (PI:
R. Ellis). The data consist of 41 single orbit pointings of 2036s
each, covering a continuous area of $19^{\prime}\times19^{\prime}$,
and was reduced according to \cite{ses07,shj09}. An extensive
description of our shape measurement pipeline is given in
\cite{shj09}. In the following we summarise the main 
characteristics. 

The shear catalogue is based on SExtractor \citep{bea96} detections,
for which we required a minimum number of 8 connected pixels with
${\rm S/N}>1.4$ each after filtering with a $5\times5$ pixel wide
Gaussian kernel. The object catalogue created in this manner was then
fed into our implementation \citep[see][]{ewb01} of the KSB method
\citep{ksb95,luk97,hfk98} for the shape measurement, adapted for
HST/ACS as detailed in \citet{ses07,shj09}. We employed a
principal component interpolation for the variable HST/ACS
point-spread function and parametric corrections for charge-transfer
inefficiency for both stars and galaxies. In addition, we applied
weights $w_i$ to the individual shear estimates given by
\begin{equation}
w_i^{-1} = \left(\frac{2}{\mathrm{Tr}[P^g_i]}\right)^2
\sigma_{e_\mathrm{ani}}^2(\mathrm{mag}_i) + 0.25^2\,,
\end{equation}
where $\left(2/\mathrm{Tr}[P^g_i]\right)$ is the isotropic PSF
correction factor for galaxy $i$ and
$\sigma_{e_\mathrm{ani}}^2(\mathrm{mag})$ denotes the variance of the
PSF anisotropy corrected galaxy polarisations fitted as a function of
magnitude. We then selected galaxies with a minimum half light radius
of $r_h>1.2\,r_h^{*,{\rm max}}$, where $r_h^{*,{\rm max}}$ is the
maximum half light radius of the 0.25 pixel wide stellar locus in a
size-magnitude diagram. An explicit magnitude cut was not
performed. For more details we refer the reader to \cite{shj09}.

After all filtering, the shear catalogue contains 33500 galaxies with
redshift estimates $z>0.3$, corresponding to a number density of
$n=73$ arcmin$^{-2}$. 42\% of the galaxies have their redshifts
estimated photometrically as outlined in Sect. \ref{photz}.

For those galaxies without redshift estimate (median magnitude 
$I_{\rm F814W}=26.0$) we used the mean magnitude-redshift relation
from \cite{shj09}. Thereto we split the galaxies into magnitude bins
of width 0.5 mag, starting from $I_{\rm F814W}=23.0$ down to 
$I_{\rm F814W}=27.5$. For each bin we calculated the average lensing
strength $\langle \beta\rangle$ defined in eq. (\ref{beta}). Since the
lens is at a very low redshift of $\zl=0.26$, it is insensitive
to the redshift distribution, in particular for galaxies with
redshifts $z\gtrsim0.7$. Essentially, $\langle \beta\rangle$ is
between 0.70 and 0.80 for 96\% of these galaxies, and we might as well
have assumed a constant redshift without affecting our results. 

The median and mean redshift of all galaxies in the shear catalogue
are 1.39 and 1.16, respectively. Objects are evenly distributed over
the sky, and the area around J0454 has only very few masks for bright
stars, none of which is larger than $\sim20^{\prime\prime}$.
We estimate the 50\% completeness limiting AB magnitude of our shear 
catalogue to $I_{\rm F814W}\sim26.1$ mag, consistent with the results
for the COSMOS field, which was observed with very similar strategies
\citep{saa07}. The depth matches the one for the ground-based data 
(see Table \ref{table_data_sup}).

\begin{figure}
\includegraphics[width=1.0\hsize]{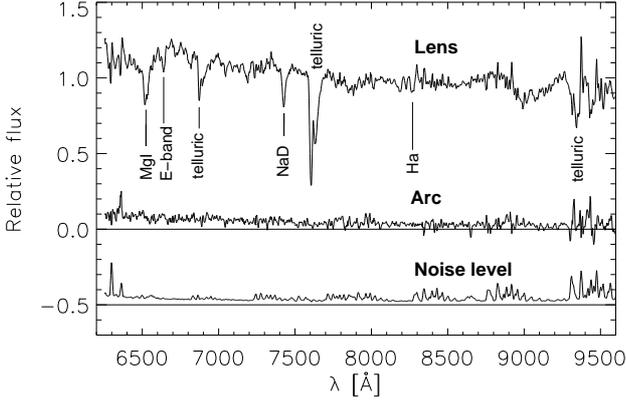}
\caption{\label{cl0454_spectra}Redshifted FORS2 spectra of the lens
  and the arc (binned 3 times). No significant features were found in
  the spectrum of the arc. The noise level was offset by -0.5 for
  better visibility.}
\end{figure}

\subsection{\label{vltfors2}VLT/FORS2 spectroscopy of the strong lens
  system}
We used VLT/FORS2 to determine the redshifts of the fossil group's
brightest elliptical, E0454, and its arc system. Data were taken on
2009-03-23 in DDT time and in 1\myarcsec0 
seeing, using the OG590 order sorting filter, GRIS\_300I grism and a
1\myarcsec0 long slit, resulting in a resolution of $R\sim660$
(1.68\angstromblank pixel$^{-1}$). The spectra were exposed for
$2\times600$s and their useable range extends over
$6250$\AA$-9300$\AA. The long slit covered the core of E0454 and the
bright northern arc. 

We debiased, flat-fielded and sky-corrected the data using modified
THELI modules. A third-order polynomial was fit to the calibration
lamp emission lines for wavelength calibration, and a small residual 
offset was corrected by comparison to sky lines. We obtained the
spectrum of E0454 by averaging 6 detector rows, yielding $S/N\sim20$
in the continuum (Fig. \ref{cl0454_spectra}). The spectrum of the arc
is strongly blended with that of E0454. To remove this contamination,
we exploited the symmetry of the lens and extracted a spectrum from
the opposite side of E0454 at the same distance as the arc. This
spectrum was subtracted from the arc's spectrum, which was then
averaged over 4 rows yielding $S/N\sim1-2$. The noise level was
determined from 200 nearby detector rows which only contained sky
background.

The lens redshift is $z=0.2594\pm0.0004$ and based on five absorption
features: MgI/MgH (5156/5196\AA), E-band (a blend 
of Fe and Ca at 5269\AA), and NaD (5890/5896\AA). Thus E0454 is a
physical member of J0454, establishing the magnitude gap and thus the
fossil character. The redshift of the arc is more difficult to
infer. Our lens modelling (see Sect. \ref{stronglens}) yields a
magnification of $8-33$ for the arc, which allows us to resolve two
maxima in its light distribution. The colours of the object and the
morphology rule out an early-type galaxy. If the morphology is
indicative of star formation and if the redshift ($z_{\rm arc}$) of
the arc is less than about $1.0$, then there would be a chance to
detect the common set of nebular emission lines such as [OII]
(3728\AA), H$\beta$ (4863\AA), [OIII] (5008\AA) and H$\alpha$
(6565\AA) with the given exposure time. However, the spectrum
does not contain any significant features. There are two possible
explanations:

First, $z_{\rm arc}$ is lower than $\sim1.0$ and the morphology
observed is not indicative for star formation, or the star formation 
rate (SFR) is low. In this case we can at least infer upper limits for
the SFR based on the non-detection of lines. For
$z_{\rm arc}=0.4$ the H$\alpha$ line would still be 
accessible. Using \cite{ken98}, a presumed line width of
$30$\angstromblank and correcting for the strong lens magnification
(see Sect. \ref{stronglens}), we find 
${\rm SFR}<0.15\,{\rm M_\odot}\,{\rm yr}^{-1}$. For 
$z_{\rm arc}=0.7-1.0$ both [OII] and H$\beta$ are covered. Following   
\cite{arl09} the upper limits for the SFR from these two lines are
${\rm SFR}<0.2-2.5\,{\rm M_\odot}\,{\rm yr}^{-1}$ for 
$z_{\rm arc}=0.7$ and ${\rm SFR}<1-10\,{\rm M_\odot}\,{\rm yr}^{-1}$
for $z_{\rm arc}=1.0$, the uncertainties being due to the unknown 
[OII]/H$\beta$ line ratio.

The second possibility is that the lensed source is at significantly
higher redshift, $z_{\rm arc}\gtrsim1.8$, such that possibly present
emission lines are redshifted beyond the spectral range covered by our
observations. The clear detection in $u^*$-band on the other hand
means $z_{\rm arc}<2.4$ and therefore $z_{\rm arc}=2.1\pm0.3$. Based
on strong-lensing properties and the stellar velocity dispersion of
E0454 we show in Sect. \ref{piemd} that this higher redshift is indeed
the most plausible assumption. The actual redshift of the arc is not
relevant for our main conclusions (see Sect. \ref{masscentroid}).

\begin{figure*}[t]
  \includegraphics[width=1.0\hsize]{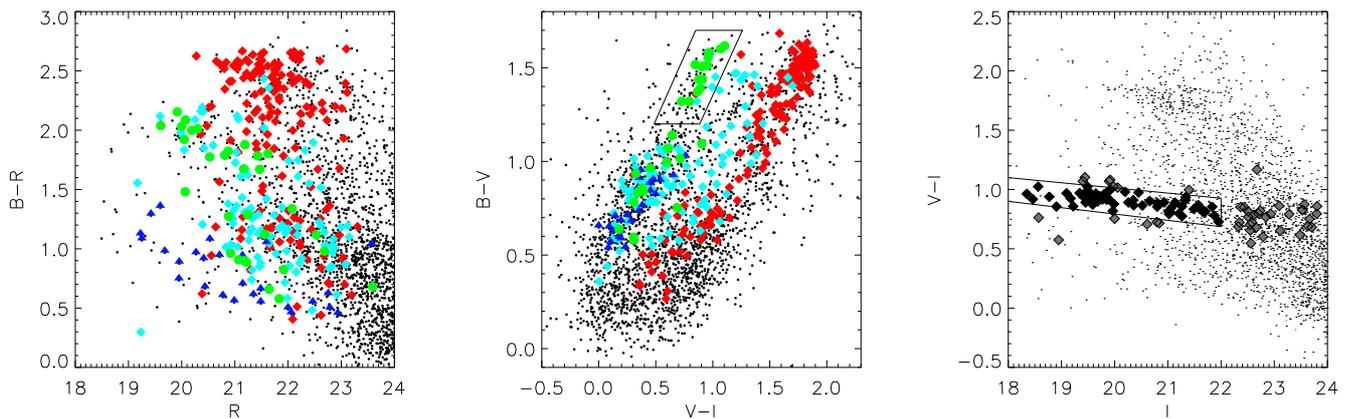}
  \caption{\label{galselection}{Target selection in colour-colour
      and colour-magnitude space. The left and middle panels: galaxies
      with $0.52<z<0.56$ and $0.28<z<0.5$ are shown as red and cyan
      diamonds, respectively. Confirmed members of J0454 are coded
      green, and blue triangles are objects with $0.1<z<0.24$. A
      selection in colour-magnitude space leads to significant
      contamination with objects at higher redshifts (left panel,
      exemplary for $B-R$ vs. $R$). Instead, we selected galaxies in
      $B-V$ vs. $V-I$ (middle). The right panel shows that the
      galaxies selected in $B-V$ vs. $V-I$ form a well-defined red
      sequence in $V-I$ vs. $I$, and the box indicates additional
      selection criteria. Black points represent galaxies that were
      kept based on this purely photometric selection, and grey ones
      were excluded. Small corrections were made by means of available
      spectra (see text for details).}}
\end{figure*}

\subsection{X-ray observations} 
The field was observed for 42 ksec on 2004-09-17 with XMM-Newton (PI:
D. Warroll, observation ID 0205670101), covering a radius of $\sim14$
around MS0451. J0454 is contained in the 2XMM catalogue \citep{wsf09}
as source 2XMM J045400.6-030832:41489. We reduced the data using
XMM-SAS\footnote{http://xmm.esac.esa.int/sas/} v8.0.0. The maximum
flare level in the $10-15$ keV range is well below 0.35 counts
s$^{-1}$, and about half of the data were taken during completely
quiescent periods. Thus we did not reject any data due to high
background rates.

X-rays are particularly absorbed by neutral hydrogen, 
\begin{equation}
I(E) = I_0\,e^{-\sigma_{\rm ph}(E)\, N_{\rm HI}}
\end{equation}
where $N_{\rm HI}=3.53\times10^{20} {\rm cm}^{-2}$ is the column
density along the line of sight \citep[taken from][]{kbh05}. Assuming
that all hydrogen atoms are in their ground state, we obtained the
quantum mechanical photon cross section as
\begin{equation}
\sigma_{\rm ph}(E) = 1.61\times10^{-23}\,{\rm cm}^2\,
  \left(\frac{E}{\rm keV}\right)^{-3.5}\,.
\end{equation}
Accordingly, the absorption is significant for low X-ray energies
($0.2-0.5$ keV) and becomes low for energies higher than $0.5-1$
keV \citep[see also][]{moc83}. For the soft cluster spectrum of J0454 
($T=1.1$ keV, see Sect. \ref{xray}), $2.5\%$ of the flux are absorbed
in the $0.5-2.0$ keV range. We also applied a k-correction factor of
1.06, interpolated from the values tabulated by \cite{bsg04}.

\section{Cluster members and field contamination}
\subsection{Object selection}
The MS0451 field had extensive wide-field spectroscopy with Keck by
\cite{met07}, who kindly made their redshift catalogue publicly
available. They randomly selected galaxies in a Subaru $I$-band
image (a subset of the data we use) from a sample with $I<21.5$ mag,
irrespective of morphology. Remaining spaces in the 14 slit masks were
then filled up with fainter objects. In total, redshifts
were obtained for 1562 galaxies in a $25^{\prime}\times20^{\prime}$
field that covers J0454 as well. 

\subsubsection{\label{memberselection}Selection of J0454 member
  galaxies}
The spectroscopic sampling of J0454 is complete to about 44\% for
$I<21.5$ mag (see Sect. \ref{contamination}). First, J0454 is at
significantly lower redshift than MS0451, and thus its brighter
galaxies were not observed as it is implausible that they are members
of MS0451. This holds in particular for the central elliptical with
$I=16.6$ mag. Second, its angular separation from MS0451 is
8\myarcminnodot and thus it was not at the centre of interest. Lastly,
slit masks cannot be configured arbitrarily due to source
clustering. 

Our photometric redshifts ($\sigma_{\rm photz}=0.040$) do not offer
sufficient power to distinguish unambiguously between
structures in the range $z=0.24-0.32$, which are present along
the line of sight (Sect. \ref{contamination}). For a more complete
picture we therefore selected ellipticals in colour-colour and
colour-magnitude space using the red cluster sequence (RCS) method
from \cite{gly00}. The spectroscopic redshifts were used to identify
suitable areas. However, the large angular extent of MS0451 and its
significant content of blue galaxies leads to a high contamination
when using the red sequence alone (see e.g. Fig. \ref{galselection},
left panel). We investigated various colour-colour combinations and
found that in $B-V$ vs. $V-I$ (Fig. \ref{galselection}, middle panel)
the highest redshift differentiation is achieved. All galaxies at
$z\sim 0.26$ are cleanly separated from those at $z=0.54$, thus
removing the bulk of the contamination. There is also very little
overlap with galaxies at $z\geq 0.3$. Only bluer objects at $z=0.26$
cannot be separated from those at lower redshift. In a first pass, we
selected objects with
\begin{equation}
B-V>1.2
\end{equation}
\begin{equation}
B-V<1.7
\end{equation}
\begin{equation}
B-V > 1.286\,(V-I) + 0.07
\end{equation}
\begin{equation}
B-V < 1.286\,(V-I) + 0.59\,.
\end{equation}
These form a red sequence in $V-I$ vs. $I$ (diamonds in the right
panel of Fig. \ref{galselection}) with a typical width of
$\sigma=0.049$ \citep[see e.g.][]{hsw09}. Only objects within 
$2\sigma$ of the red sequence and with $I\leq22$ are kept for later
analysis. The $I<22$ cut-off was chosen for two reasons. First, the
width of the red sequence increases significantly for fainter galaxies 
(see right panel of Fig. \ref{galselection}), and thus the
contamination rate would increase as well. Second, the Keck
spectroscopic survey is limited by $I\lesssim21.5$. Pushing the
photometrically selected sample significantly beyond this limit would
mean that we could not quantify anymore the contamination rate by
structures with similar redshifts.

Taken all together, these selection criteria exclude all galaxies with
$z_{\rm spec}<0.24$, and all but two ellipticals with 
$z_{\rm spec}>0.29$. A good fit to the red sequence formed by the
remaining galaxies is
\begin{equation}
V-I = -0.0430\,I + 1.768\,.
\end{equation}

From the sample of 55 galaxies selected in this manner (black
diamonds in the right panel of Fig. \ref{galselection}) we removed
three with higher and one with lower spectroscopic redshift, and
those where the photometric redshifts deviated by more than 0.1
from the cluster redshift ($2.5\sigma$ rejection, 4 objects). In
total, 47 galaxies remained to which we added 17 with confirmed
redshifts, most of them galaxies with blue colours. One galaxy 
(object \#10) had colours redder than the red sequence (caused by 
a prominent dust lane) and was added back to the sample.

In total, 15 of the red sequence galaxies have spectroscopic
redshifts, including the red galaxy that was added back to the 
sample. Assuming that similar effects hold for the 32 red sequence
galaxies without spectra, we estimate that about 2 galaxies were
overlooked. Our sample of red sequence galaxies is then $95\%$ 
complete down to $i=22.0$ mag ($M_i=-18.6\pm0.05$).

We confined the galaxy sample to within 6\myarcminnodot of the
brightest elliptical galaxy, E0454, corresponding to $1.7\,\langle
r_{200}\rangle$ (see Sects.  
\ref{r200galcounts} and \ref{tangshear}). Beyond this perimeter the
number density of red sequence galaxies is indistinguishable from 
the density of field galaxies selected in the same manner ($n=0.09$
arcmin$^{-2}$, determined from a $10^{\prime}\times11^{\prime}$ wide
area where structures with $0.2<z<0.3$ are unknown). 

\begin{figure}[t]
  \includegraphics[width=1.0\hsize]{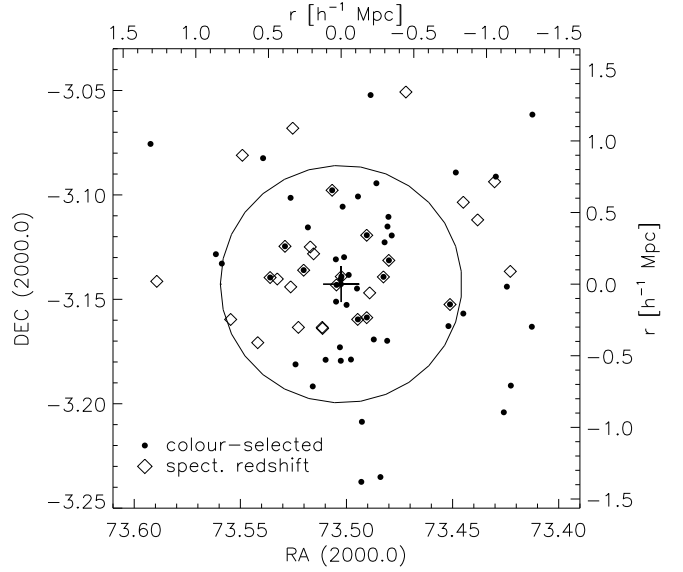}
  \caption{\label{gal_spatialdist}{Photometrically and
      spectroscopically selected cluster galaxies. The circle
      indicates $\langle r_{200}\rangle=830$ kpc, centred on E0454.}}
\end{figure}

After correcting for galactic extinction \citep{sfd98} we
determined the k-correction \citep{hbb02} using {\tt kcorrect}
\citep[v. 4.1.4,][]{blr07}. For better comparison with
other publications we report the rest-frame absolute magnitudes in the
Sloan $g$ and $i$ passbands. The errors for M$_g$ and
M$_i$ are 0.07 and 0.05 mag, respectively, based upon measurement
uncertainties and the internal error estimate of {\tt kcorrect}. The
spatial distribution of the member galaxies is shown in
Fig. \ref{gal_spatialdist}, and their properties are summarised in
Table \ref{galsample}.

\subsubsection{Magnitude gap}
Assuming a virial radius of $\langle r_{200}\rangle=840$ kpc
(3\myarcmin5) from our analysis presented below, we determined a
magnitude gap of $\Delta m_{12}=2.5$ mag in $I$-band for J0454 within
half the virial radius. The second-brightest galaxy is object \#20
from Table \ref{galsample}, an elliptical galaxy at a separation of
$0.46\,\langle r_{200}\rangle$ and with spectroscopic confirmation of
its redshift. Notice that within $0.5\,\langle r_{200}\rangle$
there is no other possible foreground or background galaxy brighter
than the second-brightest member galaxy, hence the fossil character of
J0454 is secured. The third- and fourth-brightest members within $0.5\, 
\langle r_{200}\rangle$ are 2.8 mag fainter than E0454 and also
spectroscopically confirmed. Two brighter galaxies exist at larger
radii with $\Delta_m=1.8-1.9$ (objects \#34 and \#45), but they do 
not have their redshifts measured. For a meaningful luminosity
function we need complete spectroscopic sampling, in particular
because the line of sight is contaminated by nearby structures in
redshift space (see Sect. \ref{contamination}). 

\begin{figure*}[t]
  \includegraphics[width=1.0\hsize]{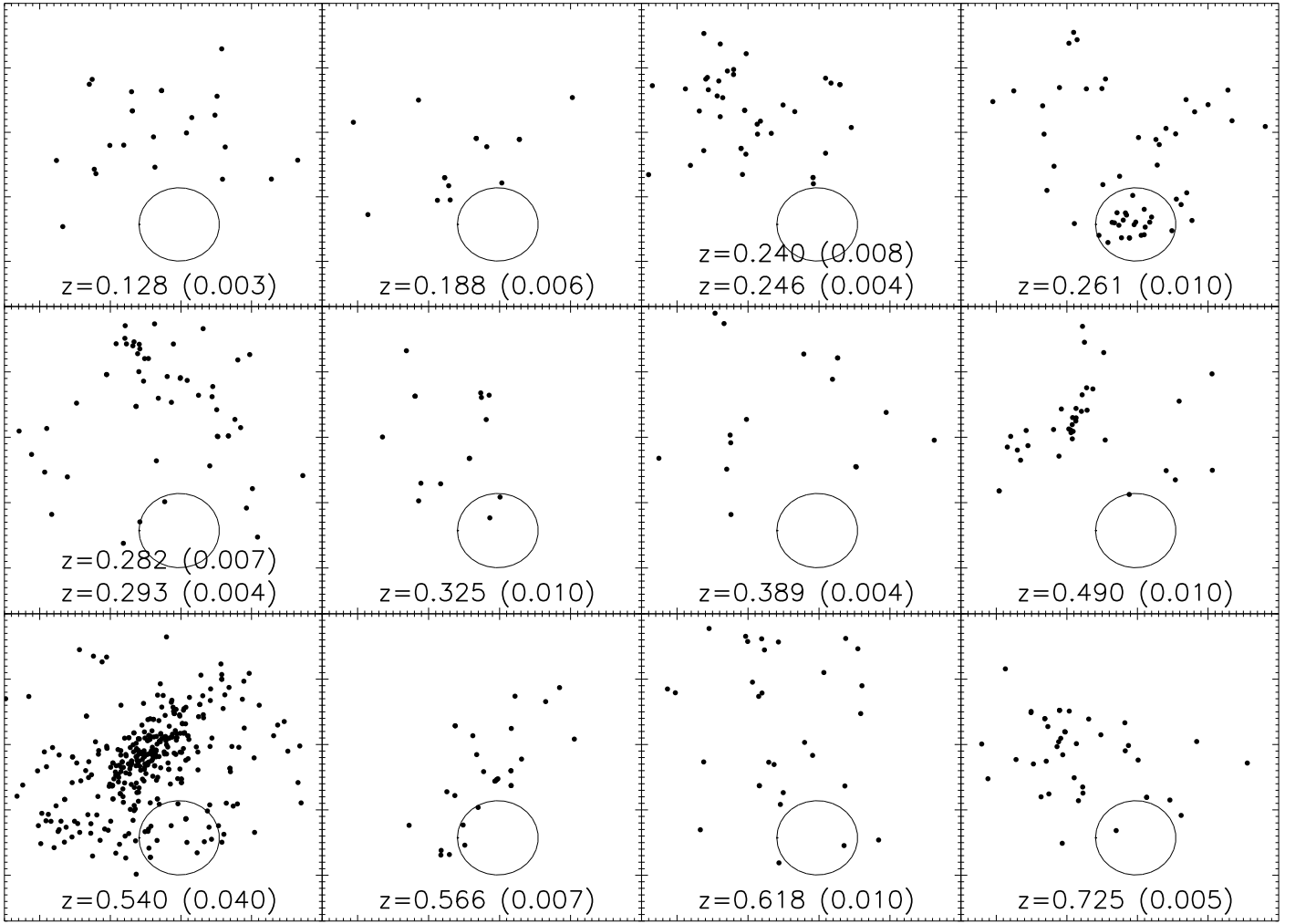}
  \caption{\label{zslices}{Clustering for different spectroscopic
      redshift bins and their width (in parentheses). J0454 is shown
      in the upper right, MS0451 in the lower left. The circle is
      centred on E0454 and traces $\langle r_{200}\rangle=830$ kpc at
      $z=0.26$. North is up and East is left. The field is $25^\prime$
      wide and centred on $\alpha=$ 04:54:06, $\delta=$ -03:02:06.}}
\end{figure*}

\subsubsection{\label{contamination}Structures along the line of sight}
Based on the Keck spectra we identified 16 structures between
$0.1<z<0.8$, consisting of at least 12 galaxies within 
$\Delta z=0.01$. The spatial distributions of the 12 most significant
ones are shown in Fig. \ref{zslices}. The circle indicates 
$\langle r_{200}\rangle=830$ kpc determined below from galaxy counts 
(Sect. \ref{r200galcounts}) and weak gravitational lensing
(Sect. \ref{tangshear}). MS0451 overlaps significantly with J0454,
whereas other structures contribute fewer interlopers.

The distributions shown in Fig. \ref{zslices} are representative of
the actual galaxy distribution. This is not self-evident due to 
the incomplete spectroscopic sampling with slit masks. However, the
main selection criterion of \cite{met07} was simply $I<21.5$ mag, with
a possible bias preferring galaxies closer to MS0451 over those with
larger separations. Thus the selection function is approximately
constant across the field and does not favour one particular structure
over another.

The line of sight towards J0454 is not only contaminated by MS0451
but also by structures at $z=0.240$, 0.246, 0.282, 0.293 and
0.325. Without spectra we cannot distinguish these from members at
$z=0.26$. We estimated the contamination assuming that the
interlopers had the same probability of being selected for
spectroscopy as the members of J0454. From the number of red 
sequence galaxies with and without spectra we determined the
spectroscopic coverage to be 44\% complete for $I<21.5$. Five
interlopers were kept by the initial selection (see Sect. 
\ref{memberselection}) and therefore we expect that about 10 of the 32 
purely photometrically selected galaxies in Table \ref{galsample} are 
not true members of J0454. We applied corrections for this where
necessary. 

\section{\label{kinematics}Morphology-density relation, kinematics and
\boldmath${r_{200}}$}
In this section we show that J0454 has characteristics
typical for normal galaxy clusters, such as a distinct
morphology-density relation \citep[see e.g.][]{gyf03} and a significantly
lower velocity dispersion for the central population of elliptical
galaxies as compared to the population of spirals. Based upon general 
cluster scaling relations, we obtain size and mass estimates.

\begin{figure}[t]
  \includegraphics[width=1.0\hsize]{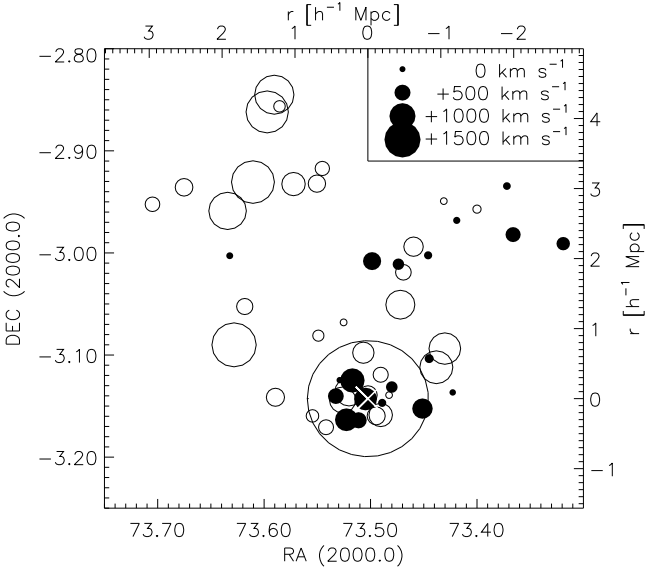}
  \caption{\label{radecz}{The kinematic structure of J0454 with
      respect to E0454. Filled and open symbols mark red- and
      blueshifted galaxies, respectively.}}
\end{figure}

\subsection{Cluster extent and mass: $r_{200}$ and $M_{200}$}
A characteristic key estimate of a cluster's linear extent is the
virial radius. It is often approximated by $r_{200}$, within which the
mean density is 200 times higher than the critical density 
$\rho_{\rm c}$,  
\begin{equation}
\label{rhocrit}
\rho_{\rm c}(z)= \frac{3}{8 \pi G} H^2(z), \;{\rm with}
\end{equation}
\begin{equation}
\label{hubble}
H^2(z) = H_0^2\;[\Omega_m(1+z)^3+\Omega_\Lambda]
\end{equation} being the Hubble function. The mass enclosed
within $r_{200}$ is
\begin{equation}
\label{m200}
M_{200} = 200\,\rho_{\rm c}(z)\,\frac{4 \pi}{3}\,r_{200}^3\,.
\end{equation}
A common estimator for the virial mass is
\begin{equation}
\label{m200vir}
M_{200}^{\rm dyn} \sim \frac{3\sigma_v^2}{G}\; r_{200}\,,
\end{equation}
which can be combined with (\ref{rhocrit}) and (\ref{m200}) yielding a
dynamic estimate for $r_{200}$,
\begin{equation}
\label{r200dyn}
r_{200}^{\rm dyn} = \frac{\sqrt{3}}{10}\frac{\sigma_v}{H(z)}\;.
\end{equation}
Estimating virial masses from galaxy dynamics is non-trivial
\citep[see e.g.][]{cye97}, in particular if the cluster under
investigation is poorly sampled with spectroscopic redshifts. Our
dynamic mass and size estimates for J0454 should therefore be viewed
with caution, and we complement them with more robust cluster scaling
relations, weak lensing and X-ray estimates.

\subsection{\label{velfield}Velocity field and virial estimate of $r_{200}$}
In Fig. \ref{radecz} we show the positions of all
galaxies around J0454 with spectroscopic redshifts in the range
$0.255<z<0.265$. The symbol size encodes the relative velocity with
respect to E0454, and open (filled) symbols denote blueshifted
(redshifted) motions. We notice two filaments extending up to
4.3 Mpc to the North and to the North-West. The former is on average
blue-shifted by $-595\kms$ compared to E0454, whereas the latter
does not show a significant motion. A photometric selection of more
member galaxies in these areas would result in significant
contamination as these filaments are projected onto four structures at
similar redshifts (Fig. \ref{zslices}). We thus confined our
subsequent analysis to the region within 6\myarcminnodot from E0454.

\begin{figure}[t]
  \includegraphics[width=1.0\hsize]{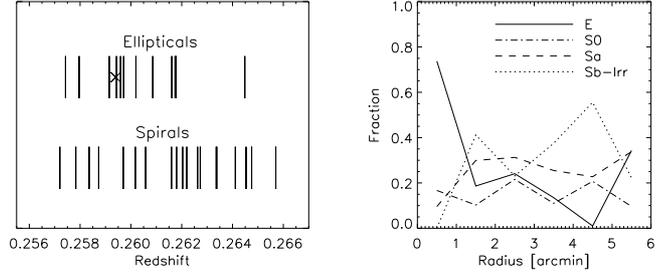}
  \caption{\label{galtypes}{Left panel: Redshift distribution for
      the red (E, S0) and blue (Sa-Sc, Irr) galaxy populations. Notice
      that the spectroscopic sampling of ellipticals is complete to
      only $\sim44\%$. The cross marks E0454. Right panel: Galaxy
      types as a function of angular separation from E0454.}} 
\end{figure}

We compute the velocity dispersion $\sigma_v$ as
\begin{equation}
\sigma_v^2=\left(\frac{c}{1+\langle z \rangle}\right)^2\,
           \left(\frac{1}{N-1}\sum_i\left[z_i-\langle
             z\rangle\right]^2 - \langle\delta\rangle^2\right)\,,
\end{equation}
excluding E0454 and following the prescription of \cite{dzt80} and
\cite{har74}. Therein, $c$ is the speed of light, $\langle z\rangle$
the mean cluster redshift, and $\langle\delta\rangle$ the uncertainty
in the redshift measurement \citep[$50\kms$, from][]{met07}. The
factor $(1+\langle z\rangle)^{-1}$ cancels the stretching effect of
cosmic expansion. After the visual classification of the galaxies'
morphologies based on their appearance in the HST/ACS data, we
determined $\sigma_v$ for the red (E, S0) and the blue (Sa-Sc, Irr)
population and for all galaxies together (see left panel of
Fig. \ref{galtypes}). Including a correction for local peculiar
motions \citep{rrs06} we have
$\sigma_v^{\rm red}=480\pm20\kms$,  
$\sigma_v^{\rm blue}=590\pm20\kms$, and 
$\sigma_v^{\rm all}=570\pm20\kms$. 
The errors were obtained from the propagated mean measurement error,
and include a conservative estimate for the uncertainty of the local
peculiar motion and a possible net motion of J0454. 

The velocity dispersion of the red galaxies is significantly lower
than the one of the blue galaxies, which is expected from dynamical
friction and the morphology-density relation \citep[right panel of Fig.
\ref{galtypes}, consistent with the findings of][for a much larger
sample of clusters]{gyf03}. Their mean velocities are different too,
and offsets exist with respect to E0454 ($+240\kms$ for the red
population, significant on the $2.5\sigma$ level, and $+540\kms$
($5.7\sigma$) for the blue galaxies). For the red galaxies this could
still be an observational effect due to incomplete sampling, as within
1\myarcminnodot of E0454 only two of nine ellipticals have their
redshifts measured. If confirmed by future observations, these
features would indicate that these galaxies have a different origin
than those which already collapsed into E0454, and that significant
substructure exists in the entire system \citep[see also][]{oeh01}.

Using equations (\ref{m200vir}) and (\ref{r200dyn}) we obtained
$r_{200}^{\rm dyn}=1054\pm44$ kpc and 
$M_{200}^{\rm dyn}=(1.69\pm0.14)\times10^{14}\,{\rm M_\odot}$ for the red
population, and $r_{200}^{\rm dyn}=1295\pm44$ kpc and 
$M_{200}^{\rm dyn}=(3.14\pm0.21)\times10^{14}\,{\rm M_\odot}$ for the blue
population.

\subsection{\label{r200galcounts}Size-richness relation}
\cite{hmw05} and \cite{jsw07} have shown that $r_{200}$ and $M_{200}$
can be estimated starting from the number $N_{\rm gal}$ of galaxies
within a radius of $1\,h_{100}^{-1}$ Mpc of the BCG. Only galaxies in
the red sequence and with $i$-band luminosities $L>0.4 L_*$ are 
considered. Based on $N_{\rm gal}$ one has
\begin{equation}
\label{hansen05}
r_{200}^{\rm gal}=0.156\, h_{100}^{-1}\; {\rm Mpc}\; N_{\rm gal}^{0.60}\,,
\end{equation}
a refined version of the original relation from \cite{hmw05}. Within
$r_{200}^{\rm gal}$ the luminosity is 200 times the mean luminosity of
the Universe. It must not be mistaken for $r_{200}$ which refers to
matter overdensity, yet the two are closely related \citep{jsw07}. 
Based on weak lensing measurements and the number $N_{200}$ of
galaxies within $r_{200}^{\rm gal}$, \cite{jsw07} and \cite{hsw09}
obtain
\begin{equation}
r_{200}=0.182\,h_{100}^{-1}\,{\rm Mpc}\,N_{200}^{0.42}
\end{equation}
\begin{equation}
M_{200}=1.75\times10^{12}\,h_{100}^{-1}\,{\rm M_\odot}\,N_{200}^{1.25}\,.
\end{equation}

Using $M_{i,*}=-21.8$ mag from \cite{hsw09}, we counted
$N_{200}=15^{+1}_{-2}$ red sequence galaxies with $M_i<-20.8$ mag
(corresponding to $0.4\,L_*$). This richness estimate contains a
correction for field contamination, and the errors are due to an
uncertainty of $0.2$ mag which we allowed for $M_{i,*}$. As a result
we have $r_{200}=811\pm46$ kpc and 
$M_{200}=(0.74\pm0.15)\times10^{14}\,{\rm M_\odot}$, including 13\%
intrinsic uncertainty for the mass-richness relation.

\section{\label{xray}X-ray halo}
The XMM-Newton image of J0454 is shown in Fig. \ref{j0454_mass_xmm},
overlaid over the HST/ACS optical image, and in Fig. 
\ref{j0454_label_cropped} in the appendix (overlaid over a colour
picture of the Subaru/Suprime-Cam data, online material). X-ray flux
is detected locally out to 1\myarcminnodot (240 kpc) from the core of
E0454, encompassing the 10 innermost galaxies. If azimuthally
averaged, we can trace the halo about twice as far. It is possible
that this very extended emission is not associated with E0454 anymore
but with the surrounding cluster of galaxies (see below). The offset
of the X-ray centroid with respect to E0454 is 
$6^{\prime\prime}\pm4^{\prime\prime}$ (24 kpc). The luminosity profile
is described by an isothermal $\beta$-model with
$\beta=0.57\pm0.06$ and a core radius of $r_{\rm c}=120\pm17$ kpc
(Fig. \ref{cl0454_xray_profile}). The best-fit isothermal redshifted
bremsstrahlung model of the spectrum yields $T=1.1\pm0.1$ keV. 
Assuming a mean particle mass of $\mu=0.6$ we find
$M_{200}=(0.34\pm0.10)\times10^{14}\,{\rm M_\odot}$ and $r_{200}=617\pm28$
kpc, respectively, and for the total luminosity within $r_{200}$ we
have $L_X=(1.4\pm0.2)\times10^{43}\,h^{-2}\ergs$. 

A cooling flow is absent from the data as can be seen from the
luminosity profile. Consequently, we do not expect star formation in
the core of E0454. This is confirmed by our VLT/FORS2 spectrum 
(Fig. \ref{cl0454_spectra}) which does not show any
H$\alpha$-emission, which would be a prime indicator for star
formation second to molecular CO emission \citep{edg01}.

The X-ray properties of J0454 agree with those of normal
groups and clusters. \cite{rmb08} find a tight correlation between
$\langle L_X\rangle$ and $\langle N_{200}\rangle$ of 17000 maxBCG
clusters, and this relation describes J0454 well. The $L_X-\sigma$
relation drawn from the same sample predicts $\sigma\sim480\pm30\kms$,
the same as we measured for the elliptical galaxy population. In the
compilation of \cite{mul00} J0454 falls comfortably within the
natural scatter of the $L_X-\sigma$ relation, resembling either a
rich group or a poor cluster.

\begin{figure}[t]
\includegraphics[width=1.0\hsize,angle=90]{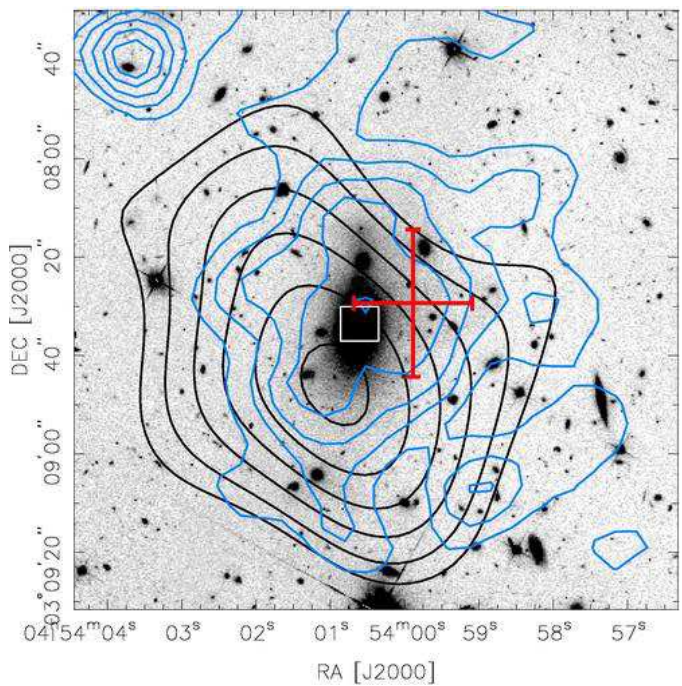}
\caption{\label{j0454_mass_xmm}HST/ACS image of J0454. The (jagged)
  blue contours trace the S/N-ratio of the $0.5-2.0$ keV X-ray flux,
  starting with $3\sigma$ and increasing in steps of $2\sigma$. A
  6\myarcsecnodot wide kernel was used for smoothing. The (smooth)
  black contours trace the S/N of the weak lensing mass reconstruction,
  starting with $2\sigma$ and increasing in steps of $0.5\sigma$,
  smoothed with a $40^{\prime\prime}$ wide kernel. The white square
  outlines the area of the strong lensing system shown in
  Fig. \ref{fig:SLImagePos}, and the red cross marks the centroid of 
  the distribution of elliptical galaxies within 
  $\langle r_{200}\rangle\sim830$ kpc.}
\end{figure}

Differences occur in temperature-based scaling relations. While no 
deviation is found with respect to the $L_X-T$ relation from the
HIFLUGCS sample \citep{seb06,reb02}, J0454 appears cooler than
expected ($\sim2$ keV) when comparing it to the $L_X-T$ relations
presented by \cite{mul00} and \cite{rmb08}. A similar trend is
seen for $T-\sigma$ \citep{mul00}, i.e. for $\sigma=480\kms$ one would
expect $T\sim2.0$ keV (or $\sigma\sim330\kms$ for $T=1.1$ keV). These
deviations can be explained by the natural scatter seen in groups of 
galaxies. A different explanation would be that we see a group-sized
substructure embedded in, but not yet fully merged with, a larger 
sparse cluster. Extended and patchy X-ray emission exists on the
lowest levels and at radii $\gtrsim 1^{\prime}$. It is unclear whether
this emission is still part of the E0454 halo or if we see the
brightest emission features of the gas associated with
J0454. With deeper X-ray data we could look for temperature variations
or different chemical compositions to distinguish these two
components. We discuss these findings in Sect. \ref{interpretation}.

The inner, flat core of the X-ray halo is elongated, tracing the
optical ellipticity of E0454. These trends have been seen
previously for groups \citep[e.g.][]{muz98} and clusters 
\citep[e.g.][]{hhb08}, and also for fossils
\citep{kjp04,sfv04,kmp06}. In general, the X-ray contours of the halo
analysed in this work are not as concentric and regular as e.g. those
for the fossil groups RX J1331.5+1108 and RX J1416.4+2315 from
\cite{kpj07}, yet they do not appear more disturbed than those of the
other three fossil groups presented by the same authors. 

We mention here that the X-ray halo of E0454 was detected
previously and is listed as object \#6 in the Chandra cluster sample
of \cite{bos02}. The reported centroid of the X-ray flux is located 
$\sim41\pm8$ kpc south-east of E0454, whereas the XMM-Newton data
reveals only a small offset of $24\pm16$ kpc to the North-West. We
explain this by the fact that XMM-Newton collected more than 10 times
as many photons as Chandra.

\begin{figure}[t]
\includegraphics[width=1.0\hsize]{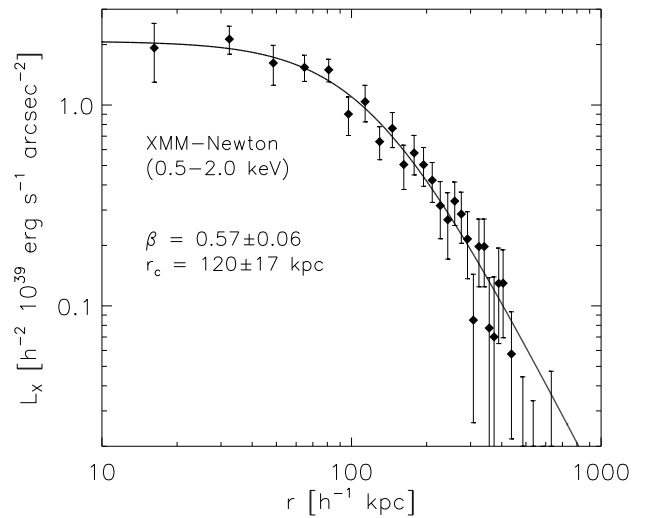}
\caption{\label{cl0454_xray_profile}Best-fit $\beta$-model for the
  X-ray halo. A cooling flow is absent.} 
\end{figure}

\section{\label{weaklens}Weak lensing analysis}
The strength of a gravitational lens scales with the ratio of the
angular diameter distances $D(z_1,z_2)$ between the lens and the source and
between the observer and the source. The more distant the source the
stronger the lensing effect, but for a lens redshift $\zl=0.26$ and 
sources at $\zs>0.8$ it is effectively constant. One must project the
sources to some arbitrarily chosen reference redshift $(\zr=1)$ and
rescale the shear estimator (the image ellipticities) accordingly to
obtain comparable shear values,
\begin{equation}
\varepsilon_{1/2}=\varepsilon_{1/2}^0 
     \frac{\D(\zl,\zr)}{\D(0,\zr)}\,
     \frac{\D(0,\zs)}{\D(\zl,\zs)}\,.
\end{equation}
This rescaling decreases (enhances) the noise for $\zs>\zr$
($\zs<\zr$) and is taken into account by individual weighting factors 
\begin{equation}
w = \left(\frac{\D(\zl,\zs)}{\D(0,\zs)} \,
          \frac{\D(0,\zr)}{\D(\zl,\zr)}\right)^2\,.
\end{equation}

Before we could proceed on the weak lensing analysis of J0454 we had
to remove the lensing contribution of MS0451 from the data by
subtracting a singular isothermal sphere (SIS) tangential shear
profile parametrised with $\sigma_v=1354\kms$. This value was taken
from \cite{cye97}, who used an iterative outlier rejection process for
its determination. A concrete error estimate was not given, but by
comparing to other measurement methods in their work, we adopted
an uncertainty of $5\%$.
Other known structures apart from MS0451 (see Fig. \ref{zslices}) do
not need to be taken into account, as their angular separation is too
large and their velocity dispersion is too low to leave a measurable
footprint at the position of J0454. The X-ray data are consistent with
this picture, revealing no structures apart from MS0451 that could add
discernible lensing signals to J0454.

\begin{figure}[t]
\includegraphics[width=1.0\hsize]{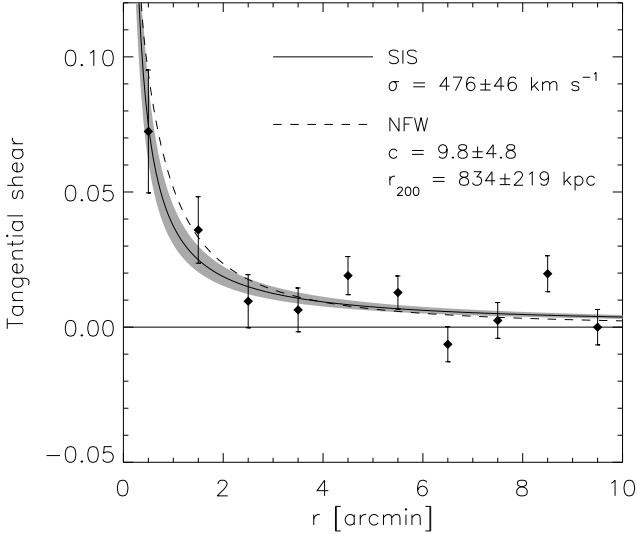}
\caption{\label{cl0454_tangshear}Tangential shear of J0454 and
  best-fit SIS and NFW profiles. The shaded area shows the 68\%
  confidence region of the SIS fit.}
\end{figure}

\subsection{\label{massrec}Mass reconstruction}
We use the finite-field method from \cite{ses01} to reconstruct the
projected surface mass density, $\kappa$, from the sheared images. 
This method uses the field border as a boundary condition, which makes
reconstructions of non-rectangular areas difficult. We therefore work
on a 16\myarcmin8 wide rectangle inscribed into the HST/ACS
mosaic. Our code is freely
available\footnote{http://www.astro.uni-bonn.de/$\sim$mischa/download/massrec.tar}  
and based on the original version from \cite{ses01}.

The convergence $\kappa$ is determined up to an additive constant, the
`mass-sheet' degeneracy, which is safely broken by assuming that
$\kappa$ vanishes on average along the border of the field. The
algorithm only works for under-critical regions with $\kappa<1$,
i.e. strong lensing areas are not reconstructed reliably. In the case
of J0454 this affects only the innermost 4\myarcsecnodot (see Sect. 
\ref{stronglens}), which is well below the resolution limit and thus
of no concern.

The resulting density map must not be interpreted without a
corresponding noise map. For example, bright stars cause holes in the
data field, which locally increase the noise due to the reduced number 
density of galaxies. In addition, the smoothing length for the shear
field must be larger than these holes. Otherwise, the boundary
condition of a rectangular data field is violated, resulting in a
corrupted solution. To obtain the noise map, we created 1000
realisations of randomised galaxy orientations keeping their positions
fixed, and obtained $\kappa$ for each. The two-dimensional rms of
these $\kappa$-maps yields the desired noise map. Since lensing
increases the ellipticities of galaxies, we removed the SIS shear
profile of J0454 (Sect. \ref{tangshear}) from the data prior to the
randomisations. Otherwise the noise at the cluster position would 
be overestimated.

The S/N-level of the mass map is shown in Fig. \ref{j0454_mass_xmm}.
J0454 is detected on the $4.7\sigma$ level with a peak convergence of
$\kappa=0.20$. It is the only significant ($S/N>4$) mass peak besides
MS0451 (${\rm S/N}=7.7$), and located $12\pm5$\myarcsecnodot south of
E0454. The uncertainty in the position was determined from
boot-strapping the shear catalogue. The mass of J0454 within 182 kpc
(approximately tracing the ${\rm S/N}=1$ contour) is
$M=(0.38\pm0.09)\times10^{14}\,{\rm M_\odot}$. This is not comparable to
$M_{200}$ since it is integrated within a much smaller radius. A
determination of $M_{200}$ from the reconstructed density map is not
sensible as the noise entirely dominates the signal in the larger
aperture. However, we can infer a lower limit of $r_{200}\gtrsim
650\pm50$ kpc. One way to test the integrity of the detection is to
check for noise peaks in the 1000 randomisations with equal or higher
significance. No such peak is found, consistent with the expectation
(0.21 peaks) from idealised Gaussian noise. In reality the noise is 
non-Gaussian as the dispersion of image ellipticities is
non-Gaussian. Probing the actual differences for $\sim5\sigma$
peaks would require many more randomisations, but would not change 
our main conclusion here that is that we detected a real signal.

As mentioned previously, we removed the contribution of MS0451 by  
subtracting a SIS profile with $\sigma_v=1354\kms$. Changing this
value by 5\% alters the mass estimate by 0.1\%, hence this measurement
is insensitive to the presence of MS0451. This is not unexpected as
the separation between J0454 and M0451 is large and $\kappa$ is a
local quantity, resulting in no overlap of the clusters' projected
surface mass densities.

\subsection{\label{tangshear}SIS and NFW fits to the tangential shear
  profile}
We fit SIS and NFW profiles to the tangential shear
around J0454 (Fig. \ref{cl0454_tangshear}), assuming a spherical
symmetric density distribution. As compared to the mass
reconstruction, the results are not model-independent. For
the SIS we furthermore assumed that the system is in virial
equilibrium with isotropic distribution of the orbits, having a
density profile 
\begin{equation}
\rho_{\rm SIS}(r)=\frac{\sigma_v^2}{2\pi\,G\,r^2}
\end{equation}
which yields, in analogy to the derivation of equation (\ref{r200dyn}),
\begin{equation}
r_{200}=\frac{\sqrt{2}}{10}\,\frac{\sigma_v}{H(z)}
\end{equation}
(note the different pre-factor).

The tangential shear is measured with respect to a reference point,
which should be near or at the centre of mass, depending on
substructure. We identify the position where the tangential shear
is maximised with a matched-filter technique \citep[the $S$-statistics
or peak finder, see][]{seh07}. The signal is maximised for a
5\myarcmin5 wide filter approximating the NFW shear profile, detecting
J0454 on the $5.2 \sigma$ level $12^{\prime\prime}\pm5^{\prime\prime}$
south-east of E0454 (position angle $162\pm2$ degrees, both error
estimates from bootstrapping). This is coincident with the peak of the
mass reconstruction and indicates a robust choice for the reference
point. In general, the two peaks would not necessarily coincide as
both methods compute very different quantities. Deviations can occur
in particular for clusters with significant substructure 
\citep[see e.g.][]{hsd09}, provided that the $S/N$ is high enough to
resolve such features. With this reference point the SIS fit yields
$\sigma_v^{\rm wl}=476\pm46\kms$ and 
$M_{200}=(0.90\pm0.26)\times10^{14} {\rm M_\odot}$, and from the NFW fit 
we obtained a concentration parameter of $c=9.5\pm4.8$, 
$r_{200}=834\pm219$ kpc and
$M_{200}=(0.84\pm0.66)\times10^{14}{\rm M_\odot}$. Contrary to the
convergence $\kappa$, the shear is a non-local quantity and therefore
more susceptible to changes in the velocity dispersion assumed for
MS0451. For example, decreasing (increasing) its $\sigma_v$ by 5\%
results in a 2\% (4\%) increase of the velocity dispersion for
J0454. These effects are included in our error budget.

To quantify the effect of possible errors in the choice of
the reference point, we repeated the analysis using the core of E0454 
and the centroid of the distribution of elliptical galaxies within
$r_{200}=830$ kpc. This yielded $\sigma_v=462\pm49\kms$ and
$376\pm58\kms$, respectively. The first fit is qualitatively slightly
worse than the original fit but still acceptable, whereas the second
is significantly deteriorated. The centroid of the distribution of
elliptical galaxies can therefore be ruled out as the centre of
mass. We show below based on strong lensing that this also
applies to E0454.

\begin{figure}
\includegraphics[width=1.0\hsize]{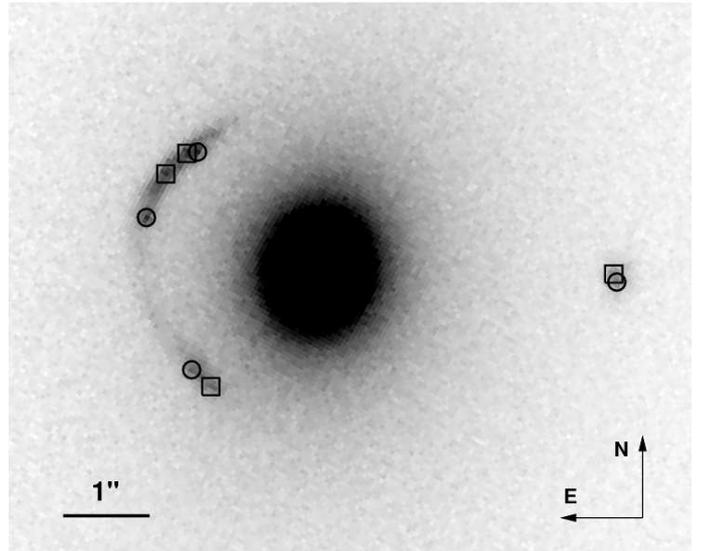}
\caption{\label{fig:SLImagePos} HST/ACS image of the strong
  lens. The counter image and the arcs reveal two maxima in the source
  intensity distribution, forming two sets of multiple image
  systems. They are marked by circles and squares and are used for 
  the lens modelling.}
\end{figure}

\section{\label{stronglens}Strong lensing analysis}
\subsection{\label{lensmodel}Lens modelling}
We identified two sets of multiple image systems, corresponding to two
bright knots in the source intensity distribution and identified by
circles and squares in Figure \ref{fig:SLImagePos}. The lens is
modelled using a pseudo-isothermal elliptic mass distribution
\citep[PIEMD,][with zero core radius]{KassiolaKovner93},
\begin{equation}
\label{eq:piemd}
\kappa(\theta_1, \theta_2) = \frac{b}{1+q}
   \left({\theta_1^2}+\frac{\theta_2^2}{q^2}\right)^{-1/2},
\end{equation}
where $b$ is the strength and $q$ is the axis ratio. The $1/(1+q)$
normalisation is needed to match the profile in
\citet{KassiolaKovner93} that was defined using ellipticities
($\epsilon \equiv (1-q)/(1+q)$) instead of axis ratios. The
distribution is translated by the centroid position and rotated by the
position angle, $PA_{\rm L}$. Furthermore, we allowed a constant
external shear with strength $\gamma_{\rm ext}$ and $PA_{\rm ext}$. In
total, there are 11 parameters: 4 for the two source positions, 5 for
the PIEMD, and 2 for the external shear. The two sets of multiple
images provide 16 constraints. Note that the modelling is independent
of lens and source redshifts.

\begin{figure}[t]
\begin{center}\includegraphics[]{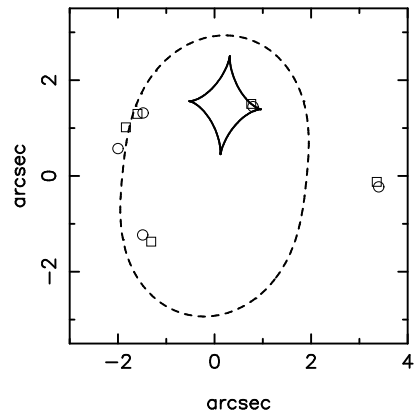}
\end{center}
\caption{\label{fig:SLCritCaus} The most probable critical curve
  (dashed) and caustic curve of the lens. The caustic
  consists of four folds (solid lines) joining at four cusps. The
  modelled source and image positions for the two sets of 
  multiple-image systems are marked by squares and circles.}
\end{figure}

We used the strong lens modelling code (Halkola et al. 2010, in
preparation) based on \citet{HalkolaEtal06}, \citet{HalkolaEtal08},
\citet{SuyuEtal06} and \citet{DunkleyEtal05}. Markov chain Monte Carlo
(MCMC) methods were employed to obtain the posterior probability
distributions of the lens parameters. We placed Gaussian priors on the
centroid, $q$, and $PA_{\rm L}$ of the PIEMD (with Gaussian widths of
$0.05''$, $0.09$ and $10^{\circ}$, respectively) based on the observed
light distribution.

Table \ref{tab:SLparams} lists the results of the marginalised lens
parameters from a MCMC chain of length $10^5$ after the burn-in phase.
Typical predicted image positions agree with the observations within 1
pixel (rms $\sim0\myarcsec03$). Figure \ref{fig:SLCritCaus} shows the
critical and caustic curves of the most probable lens parameters.
The arc is in a fold configuration, close to being in a cusp
configuration (i.e., the positions of the bright knots in the source
lie next to a fold and are in the vicinity of a cusp of the caustic
curves). The two merging images form the northern half of the arc 
with magnifications $\mu=8.2-33.8$, and the other image the southern
arc with $\mu=3.0-4.4$. The counter image has $\mu\sim2.3$.

\begin{figure}[t]
\includegraphics[width=1.0\hsize]{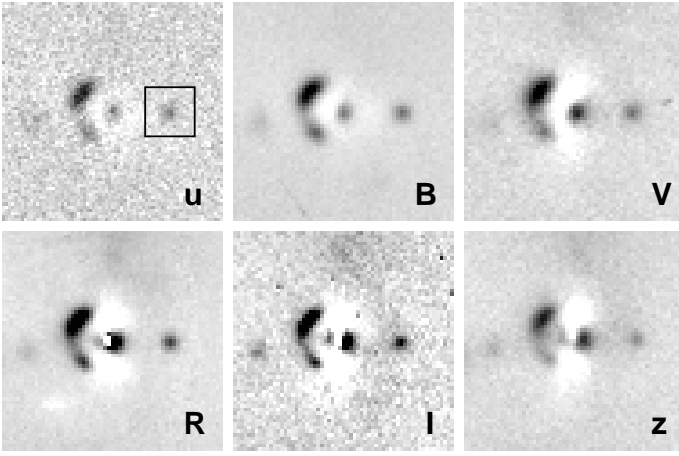}
\caption{\label{cl0454_galfit}The arc system after subtracting an
  elliptic Sersic model from the CFHT $u^*$-band and the Subaru
  $BVRIz$-band images. The lens is not entirely removed. The counter
  image is marked with a box. North is up and East is left,
  the image width is 27$^{\prime\prime}$.}
\end{figure}

\begin{figure}[t]
\includegraphics[width=1.0\hsize]{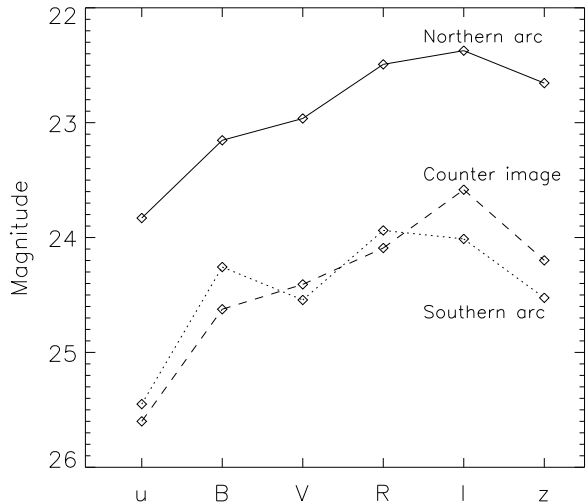}
\caption{\label{lens_colours}Magnitudes of the arc components and the
  counter image, obtained from the images in
  Fig. \ref{cl0454_galfit}. The uncertainties in the values are of the
  order of $0.2$ mag due to the residuals from the removal of
  E0454.}
\end{figure}

The separations between the arc and the lens, and between the counter
image and the lens, are $1\myarcsec85-2\myarcsec18$ and 3\myarcsec42,
respectively. This asymmetry requires the presence of significant
external shear, $\gamma_{\rm ext}=0.12$. As gravitational lensing
is an achromatic process, all images should have similar colours, 
which allowed us to test the counter image hypothesis. Since the lensed
images are very near the core of E0454 we subtracted a model for
the lens galaxy light before obtaining usable photometry. Thereto we
fit an elliptic Sersic model to the $u^*BVRIz$ data using GALFIT
\citep{phi02}. The resulting images and source fluxes are shown in
Figs. \ref{cl0454_galfit} and \ref{lens_colours}. We found good
agreement confirming the lens modelling. Only the $V$-band flux of the
southern arc appears too low, which is a consequence of the worse
seeing in this filter and the fact that this image is closest to the
lens making it very susceptible to over-subtraction effects. 

\subsection{\label{piemd}PIEMD and stellar velocity dispersions}
The equivalent Einstein radius of the PIEMD reads
\begin{equation}
\theta_{\rm E}^{\rm\,PIEMD}= 2 b\, \frac{\sqrt{q}}{1+q}
\end{equation}
\citep[e.g.][]{KoopmansEtal06} and evaluates to
$2\myarcsec37\pm0\myarcsec04$. These authors have also shown that
$\theta_{\rm E}^{\rm PIEMD}$ corresponds to that of a classical
spherically symmetric SIS, 
\begin{equation}
\theta_E=4 \pi \left(\frac{\sigma_v}{c}\right)^2 \frac{\Dls}{\Ds}\,,
\end{equation}
yielding a PIEMD velocity dispersion of 
$\sigma_v^{\rm sl}=319\pm4\kms$ for a source redshift 
of $\zs=2.1\pm0.3$.

\begin{table}[t]
\caption{Marginalised strong lens parameters. The position angles (PA)
  are counted from North to East. The uncertainties on the parameters
  correspond to the 68\% posterior credible interval.}
\label{tab:SLparams}
\begin{center}
\begin{tabular}{l c}
\hline
\hline
 $\gamma_{\rm ext}$ & $0.12\pm0.02$ \\
 $PA_{\rm ext}$ & $91_{-3}^{+5}$ [deg] \\
\hline
 $b$ &  $2.38_{-0.05}^{+0.03}$ [arcsec]\\
 $q$ & $0.80_{-0.08}^{+0.06}$\\
 $PA_{\rm L}$ & $161_{-6}^{+9}$ [deg]\\
\hline
\noalign{\smallskip}
\end{tabular}\\
\end{center}
\end{table}

\cite{elr07} provide a recipe through which the PIEMD strong 
lensing velocity dispersion can be linked to the observed stellar
velocity dispersion $\sigma^*$. Their mass model is a
parametrised truncated PIEMD (hereafter: dPIE) with core radius $a$
and scale radius $s$, which becomes identical to our model in the
limit of $a\rightarrow0$ and $s\rightarrow\infty$. We obtain
\begin{equation}
  \sigma_{\rm dPIE}=\left(\frac{2}{3}\frac{c^2}{4 \pi} 
  \frac{\Ds}{\Dls}\,\theta_{\rm E}^{\rm\,PIEMD}\right)^{1/2}= 260\pm4\kms\,.
\end{equation}
The relation between $\sigma_{\rm dPIE}$ and $\sigma^*$ is shown in
Fig. 20 of \cite{elr07}. Our PIEMD corresponds to their asymptotic
limit, approximated by the solid line in that figure. The radius $R$
($3$ FORS2 detector rows or 0\myarcsec6) within which we measured
$\sigma^*$ is significantly smaller than their effective radius $R_e$
(half mass radius), hence $R/R_e<<1$. We thus expect $\sigma^*$ to be
between $0.95\,\sigma_{\rm dPIE}$ and $1.15\,\sigma_{\rm dPIE}$ or
$250-300\kms$. A fit of the Doppler-broadened NaD absorption doublet
yields $\sigma^*=210\pm80\kms$, indicating that E0454 possibly does
not contain all the lensing mass. Up to 50\% could be located in
the halo of J0454 \citep[see also][for a similar example]{mak09}, but
the large error bars do not allow a definite conclusion.

These considerations depend on the arc redshift which
could not be determined unambiguously from spectroscopy (see
Sect. \ref{vltfors2}). For $z_{\rm arc}=0.4$ (1.0) we would have
$\sigma_{\rm dPIE}=515\,\kms$ ($345\,\kms$) and similarly high stellar
velocity dispersions. Thus $z_{\rm arc}=0.4$ is clearly ruled out by
the velocity dispersion measured ($\sigma^*=210\pm80\kms$),
and also by the fundamental plane properties of BCG galaxies
\citep{dqm07}. The latter predict $\sigma^*=280\pm35\kms$ for a BCG
with the absolute $I$-band luminosity of E0454 ($-24.1$ mag). A
redshift of $z_{\rm arc}\sim1.0$ would still be permitted within
$2\sigma$ errors, but the redshift most consistent with the data are  
$z_{\rm arc}=2.1\pm0.3$, as assumed throughout our paper.

\section{\label{interpretation}Interpretation and discussion}
\subsection{\label{masscentroid}E0454 is not at the centre of the dark
  matter potential}
Strong gravitational lensing by galaxy groups is very sensitive to the
local group environment, and in particular to the internal
distribution of dark matter \citep{mwk06}. During the build-up
of the morphology-density relation individual dark matter haloes get
partially stripped and integrated into the group halo. In these
systems, strong lensing can occur by individual galaxies with typical
Einstein radii of $\theta_E=1^{\prime\prime}-2^{\prime\prime}$, but
also by the common and more massive group halo with
$\theta_E=3^{\prime\prime}-8^{\prime\prime}$ 
\citep[see e.g.][]{fka08,lcg09}.  

The Einstein radius for the $476\pm46 \kms$ weak lensing SIS
halo of J0454 and a source redshift of $z_s=2.1\pm0.3$ is
\begin{equation}
\theta_E=4 \pi \left(\frac{\sigma_v}{c}\right)^2\,\frac{\Dls}{\Ds}
=5\myarcsec28\pm1\myarcsec02\,,
\end{equation}
more than twice as large as the observed arc radius of $2\myarcsec37$.
Hence the strong lens effect is caused by E0454 and not by the more
massive group halo. Given that all masses along the line of sight
contribute to the lensing potential, the immediate consequence of this
observation is that E0454 cannot be, on a $3\sigma$-level, located at
the centre of the gravitational potential of J0454. Otherwise the arc
radius would be significantly larger. 

This argument would not hold anymore if the lensed source was at 
very low redshift (we assume that it is at high redshift, see Sects. 
\ref{vltfors2} and \ref{piemd}). The Einstein radius
of $2\myarcsec37$ would be reproduced for $z_s=0.42$, but already for
slightly higher redshifts such as 0.5 (0.6) it would increase to
$2\myarcsec93$ ($3\myarcsec45$). But even if we overestimated the
source redshift significantly, E0454 could still not be located at
J0454's halo centre as we show in the following using the lensed image
configuration.

\begin{figure}[t]
\includegraphics[width=1.0\hsize,angle=90]{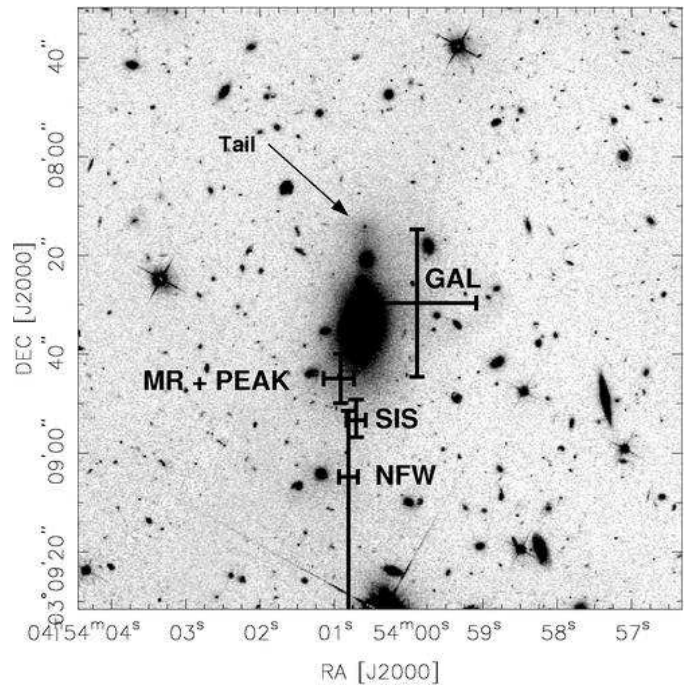}
\caption{\label{j0454_centreofmass} 
  The various estimates for the halo centre of J0454: distribution of 
  elliptical galaxies, mass reconstruction and peak finder, and from
  the external shear coming from an SIS or NFW profile. The arrow
  marks a tidal feature in the optical halo of E0454.}
\end{figure}

The lens modelling (Table \ref{tab:SLparams}) requires a large amount 
of external shear, $\gamma_{\rm ext}=0.12\pm0.02$, which can be caused
by a different lens along the line of sight, but also by an offset of
E0454 in the halo of J0454. The SIS model for the background cluster
MS0451 predicts $\gamma_{\rm ext}=0.021$ with ${\rm PA}=108$ degrees.
To obtain the external shear required, we need additional
components whose net shear is $\gamma_{\rm ext}=0.100\pm0.017$ and 
${\rm PA}=86.5\pm0.6$ degrees. Since there is no evidence for other
suitable lenses in the Keck spectra and the XMM-Newton data, this
signal must come from J0454 alone. Its centre of mass must be located
$22^{\prime\prime}\pm4^{\prime\prime}$ ($89\pm16$ kpc) south of E0454
for a SIS profile, and $31^{+30}_{-12}$ arcseconds ($126^{+122}_{-49}$
kpc) for NFW (see Fig. \ref{j0454_centreofmass}). The same
external shear could also be caused if the halo was at identical
distances to the North of E0454, but this is ruled out at the
$4\sigma$ level by the mass reconstruction and the peak finder, both
of which locate the centre of mass
$12^{\prime\prime}\pm5^{\prime\prime}$ south of E0454. This is 
unlikely to be the real centre, as neither the mass reconstruction nor
the peak finder are able to resolve such substructures in the halo 
for the given lensing signal-to-noise ratio. A much larger number
density of lensed background galaxies than $n=73$ arcmin$^{-2}$ would
be required for this purpose.

The different offsets predicted by the spherically symmetric NFW and
SIS density profiles are obviously model-dependent. Furthermore, an
elliptical halo would increase the offset if the halo's projected 
major axis pointed to E0454. The reason for this is the increased 
projected mass and thus shear seen by the strongly lensed light 
bundle towards J0454, putting the halo at larger separation from E0454
to satisfy the external shear constraint. Likewise a smaller offset
would result if the halo minor axis pointed towards E0454. The first
scenario is more likely as the distribution of red sequence galaxies
is significantly elongated North-South within $r_{200}$ (see Fig.
\ref{gal_spatialdist}), which is expected if the galaxies are
virialised within an elliptic halo. In addition, E0454 is elongated
along the same direction, and \cite{ojl09} show that the central
luminous red galaxies in clusters are preferentially aligned within
$\sim35$ degrees with their host dark matter haloes. The separations
between E0454 and the halo centre should therefore be regarded as 
lower limits.

\subsubsection{Effect of sub-haloes on the external shear}
Individual galaxies can significantly affect strong lensing systems in
cluster environments. We modelled the red sequence galaxies with SIS
profiles to estimate their contribution to the external shear, using
the velocity dispersions predicted by the Faber-Jackson relation
from \cite{dqm07}. Including successively more galaxy
haloes going from the strong lens outwards, we find that the external
shear increases gradually and stabilises at $\gamma_{\rm ext}=0.036\pm0.056$
with $PA=76$ degrees. Only 5 galaxies within 0\myarcmin51 contribute
to the signal. The uncertainty in the shear is large and based on the
intrinsic scattering of the Faber-Jackson relation. If we
systematically increase the predicted velocity dispersions of all 5
galaxies by the $1\sigma$ range allowed by Faber-Jackson, then the
entire external shear can be explained by these haloes. In the other
extreme, by lowering the velocity dispersion by $1\sigma$, the
contribution to the external shear becomes zero. 

We therefore expect halo substructures to contribute about 30\% to the
total external shear. Since this lowers the shear
coming from a smooth common group or cluster halo with 
$\sigma\sim480 \kms$, the offset from E0454 to the centre of this halo
becomes larger. For example, in case of a SIS halo the separation
would increase from 22\myarcsecnodot to 40\myarcsecnodot. The position
angle, i.e. the location of the halo centre south of E0454, remains
unchanged. We ignore the smaller effects of substructure for the rest 
of the paper as the main result, i.e. the presence of an offset of
E0454 with respect to the centre of the projected mass distribution,
remains unaffected.

\subsubsection{Interpretation: A group falling into a cluster, or a
  filament collapsing onto a group?}

How can it be explained that E0454, which is significantly more
luminous and massive than all other member galaxies, is not located 
at the minimum of the gravitational potential? Such offsets are not
uncommon for normal groups and clusters \citep{oeh01,lbk07,sby10}, 
but for old and evolved fossils they are unusual
and have not been reported previously. If all galaxies in J0454
had the same origin, then the observations are difficult to reconcile. 
A simple solution would be that E0454 formed outside of J0454 in
a separate small group which is now falling into J0454. This could
also explain the velocity offset of $+240\kms$ ($+540\kms$) observed
between the BCG and the population of elliptical (spiral) galaxies
\citep[see Sect. \ref{velfield} and][]{zam98,oeh01}. We present
more support for this scenario below based on the properties of the
X-ray halo.

There is one observational difficulty in this picture. If the halo
into which the group is falling represents a fully formed NFW mass
distribution, then one would expect a galaxy or concentration of
galaxies at its centre whose brightness is reasonably scaled to the
halo mass. However, no such galaxies are seen. From this we
conclude that the halo is not fully assembled yet, or 
decomposed into several sub-haloes in a filament projected along the
line of sight, mimicking a spherical system. In the latter case,
galaxies could be streaming along the filament onto the denser fossil
group, presenting an alternative interpretation of the system
(B. Fort, private communication). Unfortunately neither the strong nor
the weak lensing data allows us to distinguish between a filament and
a more spherical cluster. We pool both scenarios under the term
`infall hypothesis', indicating a fossil group still forming an object
of its own in a larger system (J0454).

\subsection{X-ray halo properties support infall hypothesis}
Whereas the strong lensing data require an offset of $90-120$
kpc between E0454 and the halo of J0454, only a weakly
significant offset of $24\pm16$ kpc exists between E0454 and the X-ray
halo. The latter appears to be gravitationally bound by E0454 and not
by J0454. Even though the X-ray halo overlaps in projection with the
presumed core of J0454, no significant mass transfer has happened yet
as the X-ray halo appears undisturbed. This can be explained if E0454
still forms a local minimum and thus a system of its own in the larger
gravitational potential of J0454. The significantly lower X-ray mass of 
$0.34\times10^{14}\,{\rm M_\odot}$ as compared to 
$(0.75-0.90)\times10^{14}\,{\rm M_\odot}$ from galaxy counts and weak
lensing supports this interpretation. The same holds for the
$\beta$-model velocity dispersion ($316\pm26 \kms$) which matches the
strong lensing value ($319\pm4 \kms$) much better than the one derived
from weak lensing ($476\pm46 \kms$). The halo temperature of 1.1 keV 
is also more characteristic for a group with $\sim330\kms$ than the
optically determined overall $480 \kms$. If E0454 indeed represents a
very evolved and virialised former group of galaxies that is now
falling into J0454 (Sect. \ref{masscentroid}), then one would expect
the X-ray properties to reflect a less massive and smaller system than
J0454. Given the undisturbed X-ray halo, the absence of shock fronts 
and the low temperature, we presume that E0454 has not yet passed 
through J0454 and is thus inbound for the first time. The present 
data does not allow us to infer more information about the 
three-dimensional orientation of the trajectory. To this 
end we would need a better sampling of the peculiar motions of the
member galaxies in J0454.

\subsection{X-ray offsets in other strong lensing or fossil groups}
The offset of $24\pm16$ kpc between the X-ray centroid and E0454 is
consistent with those of other groups. Four of the strong lensing
selected systems by \cite{fka08} have X-ray haloes, coinciding within
$25-50\,h_{100}^{-1}$ kpc with the brightest group galaxy (BGG). Even
smaller offsets have been observed for the five fossil groups in
\cite{kpj06}, where the X-ray centroids of four systems match those of
the BGGs. For the fifth system, RXJ1552.2+2013, an offset of 12 kpc is
reported, but the authors argue that it is unlikely to be real. Fossil
samples that were cross-matched with the ROSAT All-Sky Survey
\citep{vab99} such as those by \cite{sms07} and \cite{bcr09} show
larger offsets of up to 50 and 90 kpc, respectively. These should be
interpreted with caution though due to the poor angular resolution of
ROSAT. In general, X-ray selected galaxy groups often show small
offsets, but in dynamically disturbed or merging systems they can
become larger than 100 kpc \citep{jml06,jml07}.

Whether small X-ray offsets are representative for strong lensing
X-ray groups is currently difficult to answer due to the small
sample size and possible selection effects. For example, the groups 
from \cite{fka08} could be biased towards systems for which the X-ray
centroid and the BGG coincide, as this would boost the central density
giving rise to strong lensing. On the other hand, groups with more
complicated dynamical states (and larger X-ray offsets) have increased
lensing cross-sections and should therefore be selected as well.

\subsection{Comparison with other strong lensing groups}
\cite{lcg09} use an automated algorithm to detect strong
lensing features. They were looking for Einstein radii larger than
$3^{\prime\prime}$, targetting group-scale strong lenses, and found 13
such systems. The authors also report weak lensing measurements of the
velocity dispersions in the range of $500-800\kms$. A comparison of
the weak and strong lensing Einstein radii is not made, but the
according values have been tabulated. In general there seems to be
good agreement between the two estimates if the group halo is
responsible for the lensing. About half of the groups show
significantly larger strong lensing Einstein radii, most likely 
systems where the lensing has been boosted by the potential of an
individual galaxy in addition to the group halo. Due to the lower
cut-off in $\theta_E$ objects like J0454 with large weak and small 
strong lensing Einstein radii are filtered out. A survey aiming at
smaller strong lensing features near the core of the BGG could
identify systems similar to J0454. In combination with a high
external shear this would be a prime indicator for substructure and a
possible infall.

\subsection{Spatial and dynamic misalignment of E0454}
More evidence for the infall hypothesis arises when looking at the
sample of seven groups selected by \cite{fka08} for their strong
lensing effects. These authors found that the BGG almost always
coincides with the spatial and the dynamical group 
centre. E0454 on the other hand is marginally consistent within
$1\sigma$ with the centre of the distribution of elliptical
galaxies. In addition, its velocity deviates by $2.5\sigma$
(half the velocity dispersion) from the mean recession velocity of
the ellipticals, and even more so from that of the spiral galaxies 
(comparable to the velocity dispersion, see Sect. \ref{velfield}).
E0454 also contradicts the nine X-ray selected groups and poor
clusters of \cite{mlf06}, who found that BGGs coinciding with the
X-ray centroid have the same mean recession velocity as the
surrounding group. It would be worthwhile to look for similar
deviations in the currently existing samples of fossil groups.

With a more complete sampling of velocities of the elliptical
galaxies we could analyse these deviations in more detail, possibly
identifying a dynamic sub-population of galaxies belonging to the
fossil group (or, if we summon our alternative interpretation,
identify galaxies in the filament streaming towards the group).
With the data at hand we cannot estimate how many galaxies  
comprise the fossil group. A tidal feature in E0454's
optical halo (see Fig. \ref{j0454_centreofmass}) indicates that the
accretion process in the fossil component of J0454 has not yet
finished, and therefore it is plausible that E0454 is not the only
galaxy belonging to that component. 

\begin{table}[t]
\caption{Summary of the main results. Values in bold face are the
  primary measurements, the others were derived from these.}
\label{resultstable}
\begin{tabular}{l l l l}
\hline
\hline
\tiny
Method & $\sigma_v$ [$\kms$]& $r_{200}$ [kpc] & $M_{200}$ [$10^{14}\,{\rm M_\odot}$]\\
\hline
\noalign{\smallskip}
Galaxy counts       & --                  & \boldmath$811\pm46$   & $0.74\pm0.15$\\
Spectr. (early type)  & \boldmath$480\pm20$ & $1054\pm44$           & $1.69\pm0.14$\\
Spectr. (late type) & \boldmath$590\pm20$ & $1295\pm44$           & $3.14\pm0.21$\\
X-ray ($\beta$-model) & \boldmath$316\pm26$ & $617\pm28$          & $0.34\pm0.10$\\
Weak lens. (SIS)  & \boldmath$476\pm46$ & $853\pm82$            & $0.90\pm0.26$\\
Weak lens. (NFW)  & --                  & \boldmath$834\pm219$  & $0.84\pm0.66$\\
Weak lens. (MR)   & --                  & $650\pm50\,^*$        & \boldmath$0.38\pm0.09\,^*$\\
Strong lensing      & \boldmath$319\pm4$  & $700\pm9$             & $0.50\pm0.01$\\
\hline
\noalign{\smallskip}
\end{tabular}\\
$^*$ The mass estimate from the weak lensing mass reconstruction is
obtained within a radius of 182 kpc, significantly smaller than
$r_{200}$. The given values are therefore lower limits. 
\end{table}

\subsection{Dynamic disturbances, X-ray offsets and cooling flows}
Dynamically disturbed haloes can suppress or reheat cooling cores, as
has been shown by \cite{ses09} for the 65 systems in the Local Cluster
Substructure Survey (LoCuSS, median redshift $z=0.23$). They
demonstrated that for clusters without cooling core or with inactive
BCGs the probability distribution function of the projected offset
between the X-ray centroid and the BCG peaks between 40 and 60
kpc. Conversely, cooling core clusters never showed
offsets larger than 15 kpc. With an offset of 24 kpc and no traces of
star formation in the FORS2/VLT spectrum, E0454 matches the LoCuSS
observations. In addition there are also hints for dynamic
disturbances. E0454 is embedded in an extended optical halo (see
Fig. \ref{j0454_centreofmass}), forming a tidal tail in the North
at a distance of about 90 kpc from the core. This 
feature is too small and too localised to be attributed to a current
major merger event. It could be a residual of a recently disrupted
small companion galaxy, or caused by tidal disturbances of galaxies
orbiting very close to the halo (objects \#2, 3 and 4 from Table
\ref{galsample}). A similar feature is observed in the BCG of the
fossil cluster ESO 3060170 \citep{sfv04}.

E0454 is currently accreting mass, but the rate is too small to have
destroyed a previously existing cool core. \cite{bhg08} have
analysed the survival rate of cool cores in merger simulations and
found that non-cooling core clusters experienced a high accretion rate
with major mergers at $z>0.5$, destroying a potentially existing
cool core and also prevent their later reformation. Minor, and in
particular late accretion events such as the one observed in E0454
do not suppress a cool core. Thus, if a cooling core were present
in E0454, it must have been destroyed at early times. This is
consistent with E0454 being a giant elliptical galaxy, as only these 
galaxies have experienced major mergers in their history \citep{pef09},
and fossil systems in particular form their halos at earlier times
than other groups \citep{dkp07,bog08}. A systematic survey of fossils
regarding cooling flows and cool cores would be helpful in testing
theoretical predictions. So far, the number of examined systems is
small \citep[see e.g.][]{sfv04,kjp04,kmp06} and the amount of
fossils with suitable X-ray data poor. This also applies to J0454, for
which the data are insufficient to derive a temperature map with 
sufficient resolution to establish the non-existence of a cooling
core.

\subsection{Mass-to-light ratio}
\cite{sjm09} have obtained mass-to-light ratios for the maxBCG cluster 
sample in SDSS, using weak lensing to estimate $M_{200}$. The total
luminosity within $r_{200}$ was inferred from red sequence galaxies
only. To minimise K-corrections, it was calculated for the
$i$-band bandpass shifted to the median cluster redshift of $0.25$. We
adopt their terminology and refer to the shifted bandpass as
$^{0.25}i$. A minimum $^{0.25}i$-band luminosity of
$10^{9.5}\,h_{100}^{-2}\,{\rm L_\odot}$ was required for each galaxy.

From the SIS and NFW fit to the tangential shear profile we obtained
an average $\langle r_{200}\rangle=843$ kpc, corresponding to
3\myarcmin47. Within this radius are 14 and 22 red sequence galaxies 
with and without spectroscopic confirmation above the minimum
luminosity threshold. To correct for the field contamination
determined in Sect. \ref{contamination}, we randomly selected a
corresponding number of 5 galaxies from the sample without
spectroscopic redshifts and calculated their total flux 
contribution. This was repeated 100 times to estimate the average
background correction. The total luminosity found is
$L_{^{0.25}i}^{\rm tot}=(6.9\pm0.6)\times10^{11}\,h^{-2}{\rm L_\odot}$, and 
$M_{200}/L_{^{0.25}i}=130\pm39\,h$ for the SIS profile and 
$M_{200}/L_{^{0.25}i}=122\pm96\,h$ for NFW. In the rest-frame bandpass
the $M/L$ ratios would be 8\% lower. For a cluster with the same
number of $N_{200}$ (Sect. \ref{r200galcounts}) galaxies as J0454, 
\cite{sjm09} predict 
$\langle M_{200}/L_{^{0.25}i}\rangle=200\pm30\,h$. 
Given the scatter present in the luminosity and the mass of
a given $N_{200}$ bin \citep[see][]{sjs09,sjm09}, J0454
does not appear exceptionally over-luminous compared to non-fossil
systems. The contribution of the BGG to the total
luminosity in $i$-band within $r_{200}$ is 38\%. 
 
For completeness we also report the corresponding result for
rest-frame $B$-band and the SIS mass, $M_{200}/L_B=115\pm34\,h$. If we 
include also galaxies bluer than the red sequence, the ratio becomes
$101\pm30\,h$. The contribution of the BGG to $L_B^{\rm tot}$
is 34\% for red sequence galaxies alone, decreasing to 29\% if late
type galaxies are included. The latter is an upper limit as the sample
of late types is incomplete. For the two fossil clusters
RXJ1416.4+2315 and RXJ1552.2+2013 the non-brightest cluster members
contribute only $\sim55\%$ of the flux of the BCG, i.e. the BCG
provides about 2/3 of $L_B^{\rm tot}$
\citep[see][]{kpj07,cms06,mcs06}.

\section{Summary and conclusions}
In deep ground-based Subaru/Suprime-Cam data we discovered a
galaxy strongly lensed by a very bright elliptical galaxy
(E0454). Using VLT/FORS2 spectroscopy we confirmed that E0454 
is a member of a larger association of galaxies (J0454) at
$z=0.26$. The system forms a fossil group with a gap of 2.5 mag in
$I$-band between the brightest and the second brightest galaxy within
half the virial radius. We have spectroscopically confirmed the
membership of 31 galaxies, and furthermore selected 33 objects based
on their photometric properties. Our catalogue is complete down to
$i\leq22$ ($M_i=-18.6$). 

The data, being the deepest so far for a fossil group, show that J0454
is a complex system in various stages of mass assembly. Stripping away
the layers from outside to inside, we find two filaments extending 4
Mpc from J0454. Within a projected distance of 1.5 Mpc of the centre
is a population of spirals with $\sigma_v=590\kms$, surrounding
a more concentrated and dynamically cooler group of $\sim50$
galaxies ($\sigma_v=480\kms$). These form a red sequence with an
intrinsic width of $\sigma=0.049$.

Using HST/ACS and photometric redshifts we performed the 
first weak lensing analysis for a fossil group. The tangential shear
profile yields $r_{200}\sim840$ kpc and  
$M_{200}\sim0.85\times10^{14}\,{\rm M_\odot}$, fully consistent
with the predictions made by the cluster size-richness relation of
\cite{hsw09}. From this point of view J0454 is indistinguishable from
normal clusters, forming either a rich fossil group or a poor fossil
cluster. The X-ray halo can be described by a classic $\beta$-model
and is only marginally offset ($24$ kpc) from the brightest group
galaxy. However, the velocity dispersion of $316\kms$ is lower
than the one measured from weak lensing ($476\kms$) and
spectroscopy ($480\kms$), and so is
$M_{200}$ ($0.34\times10^{14}\,{\rm M_\odot}$). The low X-ray halo
temperature of $1.1$ keV also favours a smaller structure with 
$\sim330 \kms$.

Peculiarities arise when analysing the brightest group galaxy with
respect to its environment. It not located at the spatial centre of
elliptical galaxies and shows a significantly different velocity than
the mean velocity of the ellipticals. This indicates a different
origin of E0454 from the surrounding galaxies. More 
evidence for this hypothesis is provided by the strongly lensed galaxy
near the core of E0454. We constrained its redshift to $z=2.1\pm0.3$
and determined an Einstein radius of $2\myarcsec37$. The weak lensing
velocity dispersion of $476\kms$ corresponds to an Einstein radius of 
$\theta_E=5\myarcsec28$, meaning that E0454 cannot be located at 
the centre of the dark matter halo of J0454. Even stronger evidence
comes from the external shear required to fit the position 
of the counter image. About 15\% of the shear can be attributed
to the background cluster MS0451, and about 30\% to individual
galaxies near E0454, but the dominant contribution must come from
J0454 itself. This can only be explained if E0454 is not at 
the centre of the gravitational potential. If we describe the density
profile with NFW, then the projected distance between E0454 and the
halo centre must be at least $\sim120$ kpc. Whereas such offsets have
been shown to be common for other groups and clusters
\citep{oeh01,lbk07,sby10} this has not yet been reported for fossils.

An explanation that reconciles all observations is that E0454 is
currently infalling for the first time into the sparse cluster J0454,
seeding the brightest cluster galaxy. An alternative interpretation is
that J0454 is of filamentary nature, projected along the line
of sight, and galaxies therein stream towards the denser fossil core.
Both scenarios explain why the X-ray halo appears associated with E0454,
has undisturbed isophotes, no shock fronts, a low temperature
and a velocity dispersion and mass that fits a smaller group. This
hypothesis is only possible because of the presence and properties of
the strong lens, ruling out that E0454 is at the gravitational
centre. Without the lens all data would form a consistent picture.

Recently, \cite{lcm10} have demonstrated for the fossil UGC 842
that it segregates into two groups with $\sigma_v\sim220 \kms$ each,
separated by about $820 \kms$. Contrary to J0454 with a comparatively
low temperature of 1.1 keV, UGC 842 shows a, with respect to the
velocity dispersion, increased temperature of 1.9 keV, which has been
interpreted as a sign of an advanced interaction or merging state.

\subsection{Future observations}
Our Subaru/Suprime-Cam images are significantly deeper than those of
any other fossil system investigated so far, reaching
$10\sigma$-limits of 24.9 in $z$-band down to 26.6 in $B$-band. Hence
this data set could probe the luminosity function $\sim9$ magnitudes
below the BGG, provided deep spectroscopic data are available to
remove objects with very similar redshifts. The existing
spectra are limited to $I<21.5$ mag and 
complete to only about 40\% at this depth. Hence we limited our
analysis to galaxies with $I<22$ mag. With several hours of exposure
time at $4-8$m telescopes we could push the spectroscopic limit by
$\sim2.5$ magnitudes, which would enable us to present an
uncontaminated luminosity function extending down to dwarf
ellipticals. Numerous of those are seen in the data with colours
matching that of the red sequence. With a better spectroscopic
sampling we could remove all line of sight contamination and construct
a complete red sequence down to much fainter magnitudes, and a fairly
complete sample of blue galaxies. There are also possibilities that we 
could resolve J0454 from a dynamical point of view into members
belonging to the fossil component, and into galaxies belonging to the
sparse surrounding. If both components have indeed separate origins,
then one could attempt to identify stellar populations of different
age and composition.

Significantly deeper X-ray data could be used to better determine the
offset with respect to the BGG. We could also look for temperature
variations and changes in the chemical composition of the gas, which
would tell us more about the different origins of the sparse cluster
and the infalling group.

Lastly, deeper space-based observations could double the number
density of lensed galaxies and we could attempt to obtain direct
evidence for the separation of J0454 and E0454 in the mass
reconstructions. However, given the aged detectors of the HST/ACS 
instrument this will be a difficult endeavour.

\begin{acknowledgements}
MS thanks Bodo Ziegler at ESO and the staff at Paranal for the
prompt execution of the DDT programme, and Helen Eckmiller, Bernard
Fort, Sarah Hansen, Stefan Hilbert, Satoshi Miyazaki and Achille  
Nucita for their expertise and helpfulness concerning various aspects 
of this work. Andrew Cardwell, Karianne Holhjem and
Peter Schneider provided very useful comments on the manuscript. We
thank the anonymous referee for very helpful suggestions that improved
the paper significantly. Author contributions: MS did the
scientific analysis, obtained the VLT spectrum, reduced
the Subaru, VLT and the XMM data, discovered the strong lens system
and wrote most parts of the manuscript. SS did the strong lens
modelling, based upon a code developed by herself and by AH. TS
reduced the HST/ACS data and provided the shear catalogue, while HH
complemented it with photometric redshifts. TE provided the stacks of
the {\tt Elixir} pre-processed CFHT $u^*griz$ images. Some figures in
this paper were made with the plotting tool WIP \citep{mor95}. The
authors wish to recognize and acknowledge the very significant
cultural role and reverence that the summit of Mauna Kea has always
had within the indigenous Hawaiian community.  We are most fortunate
to have the opportunity to conduct observations from this mountain.
MS acknowledges support by the German Ministry for Science and
Education (BMBF) through DESY under the project 05AV5PDA/3 and the
Deutsche Forschungsgemeinschaft (DFG) in the frame of the
Schwerpunktprogramm SPP 1177 `Galaxy Evolution'. SS is supported in
part through the DFG under project SCHN 342/7-1. TS acknowledges
financial support from the Netherlands Organisation for Scientific
Research (NWO). HH is supported by DUEL-RTN, MRTN-CT-2006-036133, and
AH by the DFG cluster of excellence `Origin and Structure of the
Universe'.
\end{acknowledgements}

\bibliography{13810.bbl}

\begin{appendix}
\section{}
\begin{table*}
\caption{Members of J0454, ordered by increasing separation $r$ from
  E0454. An asterisk ($^*$) behind the object number indicates that it
  was not photometrically selected, but added based on its
  spectroscopic redshift. M$_g$ and M$_i$ are the absolute magnitudes
  after extinction- and k-correction. The morphological classification
  is based on the HST/ACS data and was done visually.} 
\label{galsample}
\begin{tabular}{l c c c c c c c c c c l}
\hline
\hline
\tiny
No. & $\alpha(2000.0)$ & $\delta(2000.0)$ & r [$\,^{\prime}\,$] & $I$ & $B-V$ & $V-I$ &
M$_g$ & M$_i$ & z (phot) & z (spec) & Type\\ 
\hline
\noalign{\smallskip}
1   & $04:54:00.62$ & $-03:08:33.8$ & 0.00 & 16.61 & 1.65 & 0.97 & $-23.23$ & $-24.09$ & 0.31 & 0.25940 & E\\ 
2   & $04:54:01.11$ & $-03:08:35.1$ & 0.11 & 20.65 & 1.50 & 0.91 & $-19.58$ & $-20.49$ & 0.30 & 0.26449 & S0\\
3   & $04:54:00.63$ & $-03:08:24.8$ & 0.14 & 20.80 & 1.47 & 0.85 & $-19.40$ & $-20.26$ & 0.27 & -- &E\\       
4   & $04:54:00.56$ & $-03:08:20.6$ & 0.22 & 19.64 & 1.57 & 0.97 & $-21.02$ & $-21.94$ & 0.31 & 0.25962 & E\\ 
5   & $04:53:59.74$ & $-03:08:17.9$ & 0.34 & 19.85 & 1.43 & 0.91 & $-20.75$ & $-21.54$ & 0.26 & -- &E\\
6   & $04:54:01.17$ & $-03:09:04.0$ & 0.51 & 19.89 & 1.52 & 0.99 & $-20.58$ & $-21.25$ & 0.31 & -- &S0\\      
7   & $04:53:59.98$ & $-03:09:09.6$ & 0.61 & 20.52 & 1.52 & 0.86 & $-19.70$ & $-20.40$ & 0.31 & -- &E\\       
8   & $04:54:01.19$ & $-03:07:51.0$ & 0.72 & 21.13 & 1.41 & 0.89 & $-18.63$ & $-19.45$ & 0.28 & -- &E\\       
9   & $04:54:00.26$ & $-03:07:47.3$ & 0.78 & 20.44 & 1.49 & 0.98 & $-19.30$ & $-20.24$ & 0.29 & -- &E\\      
10* & $04:53:57.36$ & $-03:08:48.7$ & 0.85 & 19.44 & 1.62 & 1.10 & $-20.96$ & $-21.99$ & 0.31 & 0.26206 & Sa\\
11  & $04:53:58.70$ & $-03:09:34.5$ & 1.12 & 20.35 & 1.37 & 0.87 & $-19.59$ & $-20.40$ & 0.26 & 0.25970 & E\\ 
12  & $04:54:04.84$ & $-03:08:09.6$ & 1.12 & 21.02 & 1.32 & 0.79 & $-18.86$ & $-19.56$ & 0.25 & 0.25796 & S0\\
13* & $04:54:03.72$ & $-03:07:41.4$ & 1.16 & 20.42 & 1.01 & 0.72 & $-19.63$ & $-20.13$ & 0.25 & 0.26272 & Sb\\
14  & $04:53:57.69$ & $-03:09:31.2$ & 1.20 & 21.06 & 1.51 & 0.85 & $-19.03$ & $-19.93$ & 0.28 & 0.25880 & SBa\\
15  & $04:53:55.81$ & $-03:08:21.2$ & 1.22 & 20.34 & 1.44 & 0.91 & $-19.62$ & $-20.52$ & 0.30 & 0.26158 & E\\ 
16* & $04:54:02.70$ & $-03:09:50.1$ & 1.37 & 21.59 & 0.64 & 0.18 & $-19.25$ & $-19.38$ & 0.22 & 0.26341 & Irr\\
17* & $04:54:04.08$ & $-03:07:29.7$ & 1.37 & 21.41 & 0.58 & 0.31 & $-18.89$ & $-19.03$ & 0.23 & 0.26476 & Sa\\
18* & $04:54:06.30$ & $-03:08:38.4$ & 1.41 & 19.40 & 1.60 & 1.07 & $-20.70$ & $-21.73$ & 0.34 & 0.25835 & Sa\\
19  & $04:53:55.20$ & $-03:07:52.8$ & 1.52 & 19.79 & 1.56 & 0.96 & $-20.10$ & $-21.09$ & 0.31 & 0.26263 & Sa\\
20  & $04:53:57.70$ & $-03:07:09.4$ & 1.58 & 19.09 & 1.58 & 0.97 & $-21.03$ & $-22.02$ & 0.31 & 0.26018 & E\\ 
21  & $04:53:55.68$ & $-03:07:21.7$ & 1.72 & 19.75 & 1.57 & 0.92 & $-20.39$ & $-21.35$ & 0.29 & -- &SBb\\    
22* & $04:54:05.43$ & $-03:09:48.3$ & 1.72 & 20.84 & 0.85 & 0.38 & $-19.58$ & $-19.85$ & 0.26 & 0.26451 & Sc\\
23* & $04:54:07.82$ & $-03:08:24.9$ & 1.79 & 21.63 & 1.14 & 0.64 & $-18.53$ & $-19.08$ & 0.32 & 0.26333 & Sc\\
24  & $04:54:00.73$ & $-03:10:22.6$ & 1.81 & 21.32 & 1.43 & 0.91 & $-18.55$ & $-19.38$ & 0.27 & -- &Sa\\     
25  & $04:53:56.91$ & $-03:10:09.2$ & 1.84 & 21.95 & 1.20 & 0.73 & $-18.01$ & $-18.72$ & 0.35 & 0.26568 & SBc\\
26  & $04:54:04.36$ & $-03:06:56.0$ & 1.87 & 21.33 & 1.39 & 0.94 & $-18.55$ & $-19.41$ & 0.26 & -- &Sb\\     
27  & $04:54:06.95$ & $-03:07:28.5$ & 1.91 & 19.56 & 1.51 & 0.95 & $-20.98$ & $-21.90$ & 0.30 & 0.26177 & S0\\
28  & $04:53:54.88$ & $-03:07:09.9$ & 2.00 & 21.66 & 1.31 & 0.81 & $-18.43$ & $-19.17$ & 0.34 & -- &Sb\\
29  & $04:54:08.62$ & $-03:08:22.3$ & 2.00 & 19.54 & 1.51 & 0.92 & $-20.47$ & $-21.40$ & 0.30 & 0.26179 & S0\\
30  & $04:53:55.39$ & $-03:10:11.5$ & 2.09 & 21.70 & 1.37 & 0.81 & $-18.14$ & $-18.94$ & 0.25 & -- & S0\\
31  & $04:53:55.36$ & $-03:06:54.4$ & 2.11 & 21.21 & 1.37 & 0.79 & $-18.83$ & $-19.56$ & 0.29 & -- & Sa\\
32  & $04:53:59.49$ & $-03:10:43.8$ & 2.18 & 20.25 & 1.37 & 0.89 & $-20.05$ & $-20.87$ & 0.29 & -- & SBb\\
33  & $04:54:00.64$ & $-03:10:45.9$ & 2.20 & 19.34 & 1.51 & 0.97 & $-20.69$ & $-21.65$ & 0.30 & -- & S0\\
34  & $04:54:02.35$ & $-03:10:44.2$ & 2.21 & 18.56 & 1.59 & 1.02 & $-21.71$ & $-22.69$ & 0.33 & -- & E\\
35  & $04:54:00.43$ & $-03:06:20.2$ & 2.22 & 20.19 & 1.52 & 0.98 & $-19.76$ & $-20.71$ & 0.31 & -- & Sa\\
36  & $04:53:55.25$ & $-03:06:37.8$ & 2.35 & 19.40 & 1.58 & 0.94 & $-20.70$ & $-21.65$ & 0.32 & -- & E\\
37  & $04:54:05.76$ & $-03:10:52.1$ & 2.63 & 21.83 & 1.32 & 0.85 & $-17.95$ & $-18.72$ & 0.22 & -- & Sa\\
38  & $04:54:01.61$ & $-03:05:51.9$ & 2.70 & 20.92 & 1.39 & 0.90 & $-18.98$ & $-19.80$ & 0.31 & 0.25915 & E\\
39  & $04:54:06.31$ & $-03:06:05.0$ & 2.85 & 21.57 & 1.23 & 0.87 & $-18.29$ & $-19.03$ & 0.26 & -- & Sb\\
40* & $04:54:10.03$ & $-03:10:14.6$ & 2.88 & 19.54 & 1.09 & 0.91 & $-20.59$ & $-21.26$ & 0.31 & 0.26021 & Sa\\
41  & $04:54:03.81$ & $-03:11:30.1$ & 3.04 & 21.48 & 1.42 & 0.84 & $-18.56$ & $-19.36$ & 0.25 & -- & SBc\\
42  & $04:53:56.61$ & $-03:05:39.9$ & 3.06 & 19.45 & 1.51 & 0.93 & $-20.74$ & $-21.62$ & 0.29 & -- & E\\
43  & $04:53:48.29$ & $-03:09:08.7$ & 3.13 & 19.98 & 1.44 & 0.88 & $-20.61$ & $-21.46$ & 0.36 & 0.26410 & SBb\\
44* & $04:54:13.09$ & $-03:09:34.7$ & 3.26 & 20.78 & 1.04 & 0.74 & $-19.51$ & $-20.15$ & 0.32 & 0.26060 & Irr\\
45  & $04:54:14.09$ & $-03:07:58.3$ & 3.40 & 18.46 & 1.60 & 0.91 & $-22.06$ & $-23.05$ & 0.30 & -- & S0\\
46  & $04:53:46.78$ & $-03:09:24.5$ & 3.56 & 19.21 & 1.48 & 0.86 & $-21.99$ & $-22.78$ & 0.27 & -- & S0\\
47  & $04:54:14.76$ & $-03:07:42.2$ & 3.62 & 21.04 & 1.34 & 0.81 & $-18.99$ & $-19.69$ & 0.25 & -- & Sa\\
48  & $04:53:58.25$ & $-03:12:31.1$ & 4.00 & 21.01 & 1.29 & 0.81 & $-19.15$ & $-19.95$ & 0.32 & -- & Sb\\
49* & $04:53:46.78$ & $-03:06:12.7$ & 4.18 & 20.87 & 0.83 & 0.40 & $-19.72$ & $-19.96$ & 0.23 & 0.26221 &Irr\\
50  & $04:54:09.42$ & $-03:04:56.6$ & 4.23 & 21.87 & 1.29 & 0.78 & $-18.08$ & $-18.71$ & 0.24 & -- & S0\\
51* & $04:53:45.16$ & $-03:06:43.0$ & 4.28 & 21.08 & 1.00 & 0.60 & $-19.84$ & $-20.21$ & 0.27 & 0.25720 & Irr\\
52  & $04:53:47.63$ & $-03:05:21.3$ & 4.56 & 19.68 & 1.44 & 0.89 & $-20.38$ & $-21.25$ & 0.28 & -- & Sa\\
53* & $04:54:11.76$ & $-03:04:51.7$ & 4.62 & 20.79 & 1.32 & 0.72 & $-19.15$ & $-19.86$ & 0.22 & 0.26083 & S0\\
54* & $04:54:06.07$ & $-03:04:05.0$ & 4.68 & 21.03 & 0.89 & 0.41 & $-19.80$ & $-20.04$ & 0.24 & 0.26156 & Sc\\
55  & $04:53:41.86$ & $-03:08:38.1$ & 4.69 & 19.78 & 1.28 & 0.92 & $-20.48$ & $-21.24$ & 0.30 & -- & SBa\\
56* & $04:53:41.47$ & $-03:08:11.8$ & 4.80 & 20.85 & 1.07 & 0.60 & $-19.50$ & $-19.97$ & 0.26 & 0.26178 & SBb\\
57* & $04:54:21.44$ & $-03:08:29.0$ & 5.19 & 20.61 & 0.94 & 0.33 & $-19.75$ & $-20.05$ & 0.22 & 0.25968 & Sb\\
58* & $04:53:43.25$ & $-03:05:37.3$ & 5.24 & 22.41 & 1.19 & 0.97 & $-17.20$ & $-18.14$ & 0.49 & 0.25743 & E\\
59  & $04:53:43.11$ & $-03:05:28.5$ & 5.35 & 19.45 & 1.47 & 0.89 & $-20.82$ & $-21.68$ & 0.28 & -- & E\\
60  & $04:53:57.27$ & $-03:03:07.8$ & 5.49 & 21.60 & 1.37 & 0.83 & $-18.20$ & $-18.97$ & 0.31 & -- & Sa\\
61  & $04:53:56.15$ & $-03:14:06.3$ & 5.65 & 19.69 & 1.46 & 0.93 & $-20.45$ & $-21.41$ & 0.32 & -- & S0\\
62  & $04:53:58.31$ & $-03:14:14.6$ & 5.71 & 19.58 & 1.44 & 0.88 & $-20.69$ & $-21.54$ & 0.29 & -- & E\\
63* & $04:53:53.27$ & $-03:03:02.5$ & 5.81 & 23.37 & 0.60 & 0.30 & $-16.84$ & $-16.97$ & 0.26 & 0.25787 & Irr\\
64  & $04:53:42.20$ & $-03:12:14.7$ & 5.89 & 19.17 & 1.58 & 0.92 & $-21.16$ & $-22.12$ & 0.30 & -- & SBa\\
\end{tabular}
\end{table*}

\begin{figure*}[t]
\includegraphics[width=0.95\hsize]{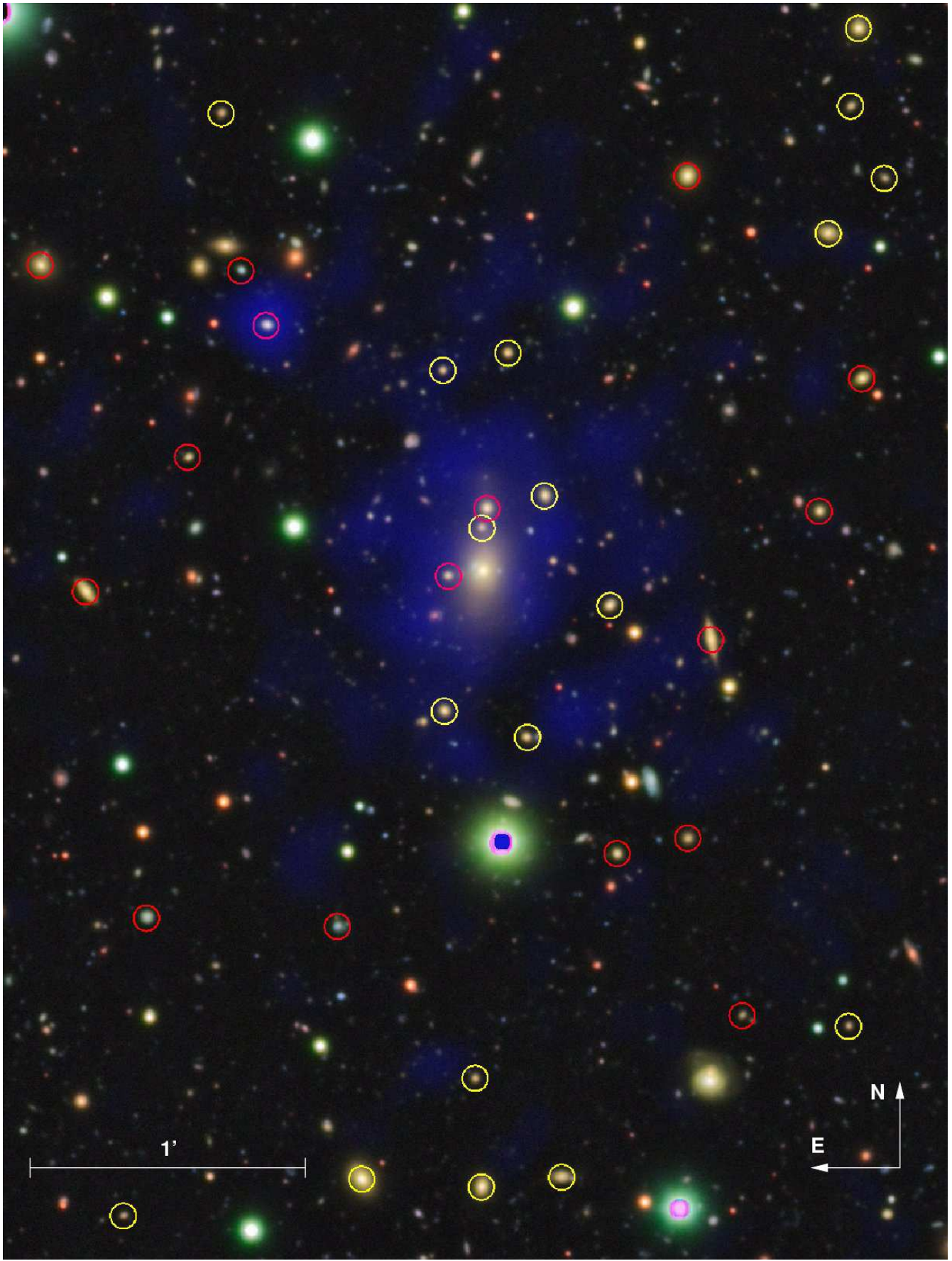}
\caption{\label{j0454_label_cropped} 
Logarithmically scaled Subaru/Suprime-Cam $BVR$ image of the central
part of J0454. Galaxies with red circles were spectroscopically
confirmed, and those with yellow circles were photometrically
selected. The brightest cluster galaxy (E0454) at the centre is left
unmarked to show the lensed system. Overlaid in blue is the (smoothed)
X-ray emission, diffuse in nature apart from the bright point source
to the upper left.}
\end{figure*}
\end{appendix}

\end{document}